\documentclass{sty/IEEEtran}

\usepackage{amsmath}
\usepackage{amsfonts}
\usepackage{amssymb}
\usepackage{graphicx}
\ifCLASSOPTIONcompsoc
\usepackage[caption=false, font=normalsize, labelfont=sf, textfont=sf]{subfig}
\else
\usepackage[caption=false, font=footnotesize]{subfig}
\fi

\usepackage{cite}

\usepackage{threeparttable}
\usepackage{multirow}
\usepackage{times}% use times font for the paper instead of default Computer Modern fonts
\usepackage{float}% better control of floating figures and tables
\usepackage{enumerate}
\usepackage[none]{hyphenat}% turn off hyphenation to make text extraction and indexing easier
\usepackage[normalem]{ulem}
\usepackage{color}
\usepackage{xcolor}

\usepackage{tikz}
\usepackage{pgfplots}
\pgfplotsset{compat=newest}
\usetikzlibrary{positioning}

\usepackage{algorithm}
\usepackage{algorithmic}

\newtheorem{remark}{Remark}

\newcommand{\qed}{\nobreak \ifvmode \relax \else
	\ifdim\lastskip<1.5em \hskip-\lastskip
	\hskip22em plus0em minus0.5em \fi \nobreak
	\vrule height0.4em width0.3em depth0.25em\fi}

\newlength\fheight
\newlength\fwidth
\begin{document}
\title{Narrowband Electromagnetic Coupling Matrix in Coupled-Resonator Microwave Circuits}

%
% \author{Valent\'{i}n de la Rubia
% 	\thanks{Valent\'{i}n de la Rubia is with the Departamento de Matem\'{a}tica Aplicada a las {TIC}, ETSI de Telecomunicaci\'{o}n, Universidad Polit\'{e}cnica de 
% 	Madrid, 28040 Madrid, Spain (e-mail: valentin.delarubia@upm.es).}
% }
%
\author{Valent\'{i}n de la Rubia and David Young
	\thanks{Valent\'{i}n de la Rubia is with the Departamento de Matem\'{a}tica Aplicada a las {TIC}, ETSI de Telecomunicaci\'{o}n, Universidad Polit\'{e}cnica de 
	Madrid, 28040 Madrid, Spain (e-mail: valentin.delarubia@upm.es).}
	\thanks{David Young is with Huawei Technologies Sweden AB, 164 94 Kista, Sweden (e-mail: david.young@huawei.com).}	
}
%
% \author{Valent\'{i}n de la Rubia, Simone Bastioli~\IEEEmembership{Senior Member,~IEEE} and Richard V. Snyder~\IEEEmembership{Life Fellow,~IEEE}
% 	\thanks{Valent\'{i}n de la Rubia is with the Departamento de Matem\'{a}tica Aplicada a las {TIC}, ETSI de Telecomunicaci\'{o}n, Universidad Polit\'{e}cnica de 
% 	Madrid, 28040 Madrid, Spain (e-mail: valentin.delarubia@upm.es).}
% 	\thanks{Simone Bastioli and Richard V. Snyder are with RS Microwave Company Inc., Butler, NJ 07405, USA (e-mail: sbastioli@rsmicro.com; r.snyder@ieee.org).}
% }
%
\maketitle
\begin{abstract}
	A novel methodology to unleash electromagnetic coupling matrix information in coupled-resonator microwave circuits has been recently proposed \cite{delaRubia2022EMCouplingMatrix}. This information is derived from Maxwell's equations and the natural language of electromagnetics is employed. As a result, the coupling matrix coefficients stand \emph{only} for electromagnetics. In this work, we enhance this approach to reveal valuable design information for microwave engineering, showing the electromagnetic (EM) coupling \emph{among all} EM resonators and ports. By the same token, the similarities with the well-known classical coupling matrix theory are addressed. We bridge this gap since the classical theory is the preferential language among microwave engineers. %Surprisingly, this latter natural language is more artificial, since classical coupling matrix theory, which stands upon Kirchhoff's laws, is in fact a narrowband approximation to electromagnetics.

	% A narrowband approximation to the electromagnetic coupling matrix is carried out since classical coupling matrix theory is a narrowband model for electromagnetics. 
	Classical coupling matrix theory is a narrowband model for electromagnetics. Thus, we carry out a narrowband approximation in the electromagnetic coupling matrix. This makes it possible to describe the EM coupling coefficients in the same framework as classical circuits. As a result, proper comparison between both coupling matrices is allowed.
	
	Finally, both coupling matrix approaches {\color{black}have} a common {\color{black}ground}, namely, get physical insight and valuable information for design purposes in coupled-resonator microwave circuits. However, \emph{only} the electromagnetic coupling matrix {\color{black}details} all EM behavior, including {\color{black}parasitic and leakage couplings, and} the higher-order mode influence in the microwave circuit. Several microwave circuits, such as filters and diplexers, will show the possibilities of this new technique and its relation to classical coupling matrix theory.
\end{abstract}
%
% \markboth{IEEE Transactions on Microwave Theory and Techniques}{de la Rubia and Young: Narrowband Electromagnetic Coupling Matrix in Coupled-Resonator Microwave Circuits}
% \markboth{IEEE Transactions on Microwave Theory and Techniques}{de la Rubia: Narrowband Electromagnetic Coupling Matrix in Coupled-Resonator Microwave Circuits}
%
\IEEEpeerreviewmaketitle
\section{Introduction}
\label{Sec-Introduction}
\PARstart{C}{ommunication} systems are in full plethora urging microwave engineers to {\color{black}rely} on time-consuming simulations in order to carry out robust electrical designs, which need to satisfy stringent electrical specifications. {\color{black}As a result of this complexity}, traditional design techniques, typically based on circuit approximations, can not {\color{black}fully accomplish} the target electrical design and full-wave optimization becomes a must in any industrial design. All EM phenomena are taken into account in these full-wave approaches to carry out accurate designs. However, each single full-wave simulation can be rather time-consuming, never mind considering an optimization loop, where a large amount of simulations needs to be accomplished. From the design engineer perspective, wouldn't it be beneficial to get all design information by means of one single full-wave simulation? {\color{black}This work is a step forward along this direction.} %This work is a clear step forward along this direction. %How are we going to do it? Are we there yet?
%circuit approximation approaches can not complete the target electrical design and full-wave optimization seems to be a must. All EM phenomena are taken into account in these full-wave approaches to carry out accurate designs. Each single full-wave simulation can be rather time-consuming, never mind considering an optimization loop, where a large amount of simulations needs to be accomplished. From the design engineer perspective, wouldn't it be beneficial to get all design information by means of one single full-wave simulation? How are we going to do it? Are we there yet?
 
Numerical codes in computational electromagnetics (CEM) predict output data to input responses and, as a result, {\color{black}we \emph{only} get} transfer function information. Little knowledge of the underlying physics in the system is actually provided \emph{so far}. This is the rationale why we need to rely on slowly convergent full-wave optimization loops (large number of iterations until we explore the design parameter space) to carry out an EM design. After all, we are still far away from getting design information out of one single full-wave simulation. However, there is an increasing need in industry to come up with smarter EM design strategies. In this work, we dive into these new design paradigms. As a matter of fact, different approaches are proposed in the literature to extract circuit information out of simulations as well as measurements \cite{macchiarella2006formulation,macchiarella2007robust,lin2007vectorfitting,vahldieck2001automated,zaki2002computer,sabry2002computer,meng2006analyticaldiagnosis,ziegler2016coupling,michalski2021coupling,zhao2018circuit,lamecki2004fast,meng2009analytical,macchiarella2010extraction,hu2014generalized,zhao2016model,zhao2017adaptive,peng2010parameter,zhao2016new,oldoni2023analyticalderivation,wu2023agenerelizedcircuitmodel,ming2023circuitmodel,ming2023filtersingeneralized,ming2023novelcompact,rodriguez2012analytical,boix2015simpleandcompact,mesa2018unlocking,rodriguez2018resonant,mouris2020increment}. This is a clear attempt to bring some design information into the full-wave simulation. To the best of our knowledge, this physical insight has never been provided by any CEM code \emph{itself}. This has been the general framework until the recently proposed approach in \cite{delaRubia2022EMCouplingMatrix}.

In this work, we deal with an enhanced methodology for EM design, which can be applied to any coupled-resonator microwave circuit. Our starting point is the impedance matrix response as a function of frequency detailing the EM behavior of the coupled-resonator EM circuit. This impedance matrix frequency response, written down in pole-residue form, is obtained with ease by means of a full-wave simulation of the microwave circuit. As a matter of fact, we deal with the numerical solution to Maxwell's equations, namely, the finite element method (FEM) and a model order reduction (MOR) layer to speed the computation of the frequency response \cite{delarubia2009reliable,delaRubia2018CRBM,morHesB13,delaRubia2014Reliable,morFenB19,hess2015estimating,Silveira2014Reduced-OrderModels,Edlinger2017finite,Edlinger2014APosteriori,Edlinger2015ANewMethod,nicolini2019model,Rewienski2016greedy,MonjedelaRubia2020EFIE,codecasa2019exploiting,xue2020rapid,szypulski2020SSMMM,Kouki,feng2019new,Vouvakis2011FastFrequency,Edlinger2017finite,chellappa2023infsup,delaRubia2022physicsbased,delaRubia2022fastaposteriori,fotyga2021amodelorder,fotyga2022MTT} are used to get this pole-residue representation of the impedance matrix in electromagnetics.
%
%We start out from the impedance matrix in pole-residue form in electromagnetics. This impedance matrix frequency response is obtained with ease by means of a full-wave simulation of the microwave filter. As a matter of fact, we deal with the numerical solution to Maxwell's equations and the FEM and a MOR layer to speed the computation of frequency response are used to get this pole-residue representation of the impedance matrix in electromagnetics.

A dynamical system with valuable design information can be {\color{black}obtained} out of the frequency response in pole-residue form of the impedance matrix in electromagnetics. {\color{black}This will be discussed in detail in Section \ref{Sec-ImpedanceMatrix}}. This new dynamical system representation can, to some extent (in fact in the narrowband limit), be identified to a circuit description of the coupled-resonator EM circuit. However, we do not claim the new dynamical system representation is a circuit at all, because it is not; but, this dynamical system somehow provides valuable EM coupling information. We will further elaborate on this concept and show how much physical insight of the microwave circuit can be obtained by means of one single full-wave simulation. As a result, we are able to get an electromagnetically generated coupling matrix, i.e., an \emph{electromagnetic coupling} matrix, detailing the underlying physics in the coupled-resonator EM circuit at the computational cost of only one single full-wave simulation.

We should point out that this new microwave circuit dynamical system representation is not related to the well-known circuit theory coupling matrix at first glance, so no similar behavior should be expected. For instance, narrowband limitation is no longer a constraint within the new methodology. The coupling matrix from circuit theory is a completely different concept since it does not stand for electromagnetics, i.e., it is used as an approximate model for electromagnetics. \emph{Only} the electromagnetic coupling matrix holds all EM behavior, including {\color{black} the parasitic and leakage couplings, and} the higher-order mode loadings in the coupled-resonator EM circuit. In a first sight, {\color{black}\emph{we kindly ask the readers to dissociate these two concepts}. As such, there is no need to remove any phase loading effect to get the electromagnetic coupling matrix, for instance.} However, also in this work, we bridge the gap between these two coupling matrices since we are aware that the classical coupling matrix is an established and well-known methodology in microwave engineering. This bridge is a must to proselytize this new EM approach. As a result, a narrowband approximation in the electromagnetic coupling matrix is detailed to properly compare both approaches.

This work is organized as follows. In Section \ref{Sec-ImpedanceMatrix}, we review the impedance matrix transfer function which details all EM behavior in the microwave circuit following Kurokawa theory \cite{Kurokawa,Conciauro,Kirsch,delaRubia2018CRBM} and identify the electromagnetic coupling matrix{\color{black}\cite{delaRubia2022EMCouplingMatrix}}. By the same token, the classical circuit theory coupling matrix is reviewed. Section \ref{Sec-NarrowbandAnalysis} discusses in detail the narrowband approximation in the electromagnetic coupling matrix to turn the dynamical system arisen in electromagnetics into a circuit theory dynamical system along a specific frequency interval (so-called narrowband analysis). In Section~\ref{Sec-NumericalResults}, we analyze several coupled-resonator microwave circuits, such as a diplexer and different type of filters, to illustrate the capabilities and accuracy of the proposed methodology. Finally, in Section \ref{Sec-Conclusions}, we comment on the conclusions.
\section{Impedance Matrix Transfer Function}
\label{Sec-ImpedanceMatrix}
%
% Our starting point is the impedance matrix of the microwave filter. 
The frequency response of the impedance matrix completely describes the EM behavior of a microwave circuit. Contrary to the classical behavior in circuit theory, in a two-port microwave circuit, the $2 \times 2$ impedance matrix has \emph{infinitely many} contributions in frequency in the \emph{electromagnetic spectrum}, as shown in \cite{Kurokawa,Conciauro,amari2007physical,amari2010theory}; each of them corresponding to the (\emph{infinitely many}) eigenmodes present in the microwave circuit \cite{Kirsch,delaRubia2018CRBM,delaRubia2022EMCouplingMatrix}. This gives rise to a Kurokawa series expansion of the impedance matrix in electromagnetics, which is in fact a pole-residue series representation of this impedance matrix transfer function, viz.
\begin{equation}
    \label{eq:Sec-ImpedanceMatrix-ImpendanceMatrix}
    \begin{aligned}
    \begin{pmatrix}
    v_1 \\
    v_2 
    \end{pmatrix} &= j k \eta_0 \sum \limits_{n=0}^{\infty} \frac{ \begin{pmatrix} c_{n1} \\ c_{n2} \end{pmatrix}   \begin{pmatrix} c_{n1} & c_{n2} \end{pmatrix} }{k_n^2 - k^2 } \begin{pmatrix}
    i_1 \\
    i_2
    \end{pmatrix}\\
    \mathbf{v} &= \mathbf{Z}(k) \mathbf{i}\text{.}
    \end{aligned}
\end{equation}
$k_n$ are the eigenresonances corresponding to each eigenmode (accumulating at infinity and related to the poles) and $c_{np}$, with $p=1,2$, are the coupling coefficients of each eigenmode to the corresponding port $p$ (related to the residues). $v_p$~and~$i_p$ are the voltages and currents on port $p$. Finally, $j$ denotes the imaginary unit, $k$ stands for the wavenumber and $\eta_0$ is the intrinsic impedance in vacuum. $\mathbf{Z}(k)$ is the impedance matrix transfer function. {\color{black}We use the term
frequency for both the wavenumber $k$ and the frequency itself.} It should be pointed out that each pole-residue term in the series \eqref{eq:Sec-ImpedanceMatrix-ImpendanceMatrix} has valuable EM information. As a matter of fact, the pole is the square of the resonance frequency of the corresponding eigenmode, and the associated matrix residue for each pole details the coupling coefficient from that eigenmode to port $p$. See \cite{Kurokawa,Conciauro,bekheit2008modeling,delaRubia2022EMCouplingMatrix} for all the details. {\color{black}Note that the matrix residues in \eqref{eq:Sec-ImpedanceMatrix-ImpendanceMatrix} must be rank-1 matrices, cf. \cite{delaRubia2022EMCouplingMatrix,delaRubia2022PhysicsBasedGreedyAlgorithm}.}

Even though \emph{infinitely many} eigenmodes are present \emph{in electromagnetics}, just a few of them are responsible for the dominant electrical behavior of the microwave circuit in the band of interest, due to its finite bandwidth. This gives rise to the concept of the order of the coupled-resonator microwave circuit, representing the number of dominant resonators in the circuit. These are the ones which resonances arise in the band of interest, so-called in-band eigenresonances (\emph{finitely many}), {\color{black}and stand for the dominant behavior of the coupled-resonator microwave circuit}. However, if we still wish to keep ourselves under the {\color{black}umbrella} of electromagnetics, the out-of-band eigenresonances must not be neglected. These are the (\emph{infinitely many}) higher-order modes, along with statics, arisen \emph{in electromagnetics}. As a result, the impedance matrix $\mathbf{Z}(k)$ is in fact \emph{twofold}: the in-band eigenresonance part $\mathbf{Z}_\text{in-band}(k)$, {\color{black}standing for the dominant EM behavior in the band of interest}, and the out-of-band eigenresonance contribution $\mathbf{Z}_\text{out-of-band}(k)$, {\color{black}typically contributing with a phase loading effect at the ports}. Any CEM code should be able to provide this impedance matrix transfer function in the band of interest and its in-band/out-of-band contributions in a clear way. In our case, we use the FEM and a MOR layer on top of it for fast frequency sweep computation {\color{black}to obtain this result with ease} \cite{delarubia2009reliable,delaRubia2014Reliable,delaRubia2022spurious,delaRubia2022physicsbased,delaRubia2022fastaposteriori,fotyga2022MTT,chellappa2023infsup,fotyga2022MTT,delaRubia2018CRBM,szypulski2020SSMMM,chellappa2021adaptive}. As a result, \eqref{eq:Sec-ImpedanceMatrix-ImpendanceMatrix} can be written down in a more compact form, viz.
\begin{equation}
    \label{eq:Sec-ImpedanceMatrix-ImpendanceMatrix2}
    \begin{aligned}
		\mathbf{v} &= \mathbf{Z}(k) \mathbf{i} =  (\mathbf{Z}_\text{in-band}(k) + \mathbf{Z}_\text{out-of-band}(k) )\mathbf{i} \\
		&= \mathbf{v}_\text{in-band}+\mathbf{v}_\text{out-band} =  \mathbf{Z}_\text{in-band}(k) \mathbf{i} + \mathbf{Z}_\text{out-of-band}(k) \mathbf{i} \\
	\begin{pmatrix}
    v_1 \\
    v_2 
    \end{pmatrix} &= j k \eta_0 \sum \limits_{n=1}^{N} \frac{ \begin{pmatrix} c_{n1} \\ c_{n2} \end{pmatrix}   \begin{pmatrix} c_{n1} & c_{n2} \end{pmatrix} }{k_n^2 - k^2 } \begin{pmatrix}
    i_1 \\
    i_2
    \end{pmatrix} \\
    &+ \mathbf{Z}_\text{out-of-band}(k) \begin{pmatrix}
    i_1 \\
    i_2
    \end{pmatrix} \text{.}
    \end{aligned}
\end{equation}
$N$ is the order of the coupled-resonator microwave circuit and $k_1,\dots,k_N$ stand for the in-band eigenresonances responsible for the dominant EM behavior in the band of interest, namely, $\mathbf{Z}_\text{in-band}(k)$. Note that $\mathbf{v}$ has also been split {\color{black}in \eqref{eq:Sec-ImpedanceMatrix-ImpendanceMatrix2}} into these in-band/out-of-band contributions, i.e., $\mathbf{v}_\text{in-band}$ and $\mathbf{v}_\text{out-of-band}$. 

{\color{black}For the sake of understanding, we detail the expressions for all, the impedance matrix transfer function $\mathbf{Z}(k)$, and the in-band and out-of-band impedance matrix transfer functions $\mathbf{Z}_\text{in-band}(k)$, $\mathbf{Z}_\text{out-of-band}(k)$, viz.
\begin{subequations}
    \label{eq:Sec-ImpedanceMatrix-SeriesDefinition}
    \begin{align}
		\label{eq:Sec-ImpedanceMatrix-SeriesDefinitionAll}
		\mathbf{Z}(k) &= j k \eta_0 \sum \limits_{n=0}^{\infty} \frac{ \begin{pmatrix} c_{n1} \\ c_{n2} \end{pmatrix}   \begin{pmatrix} c_{n1} & c_{n2} \end{pmatrix} }{k_n^2 - k^2 }\\
		\label{eq:Sec-ImpedanceMatrix-SeriesDefinitionInBand}
		\mathbf{Z}_\text{in-band}(k) &= j k \eta_0 \sum \limits_{n=1}^{N} \frac{ \begin{pmatrix} c_{n1} \\ c_{n2} \end{pmatrix}   \begin{pmatrix} c_{n1} & c_{n2} \end{pmatrix} }{k_n^2 - k^2 }\\
		\label{eq:Sec-ImpedanceMatrix-SeriesDefinitionOutOfBand}
		\mathbf{Z}_\text{out-of-band}(k) &= j k \eta_0 \sum \limits_{\substack{n=0 \\ n \notin \{1,\dots,N\}}}^{\infty} \frac{ \begin{pmatrix} c_{n1} \\ c_{n2} \end{pmatrix}   \begin{pmatrix} c_{n1} & c_{n2} \end{pmatrix} }{k_n^2 - k^2 }\text{.}
    \end{align}
\end{subequations}

Now, we focus on the in-band dominant eigenresonances in the impedance matrix transfer function, $\mathbf{Z}_\text{in-band}(k)$, detailed in \eqref{eq:Sec-ImpedanceMatrix-SeriesDefinitionInBand} and arisen in \eqref{eq:Sec-ImpedanceMatrix-ImpendanceMatrix2} as well.} These dominant pole-residue contributions can {\color{black}be actually cast} into matrix form to yield a dynamical system representation, viz. %for $\mathbf{Z}_\text{in-band}(k)$, viz.
\begin{subequations}
	\label{eq:Sec-ImpedanceMatrix-DynamicalSystem}
	\begin{align}
	\label{eq:Sec-ImpedanceMatrix-DynamicalSystem1}
	\mathbf{v}_\text{in-band} = j k \eta_0 \mathbf{C} (\mathbf{K}-k^2\mathbf{Id})^{-1} \mathbf{C}^T \mathbf{i} &= \mathbf{Z}_\text{in-band}(k) \mathbf{i}
	\\
	\label{eq:Sec-ImpedanceMatrix-DynamicalSystem2}
	\begin{pmatrix}
	\mathbf{0} & \mathbf{C} \\
	\mathbf{C}^T & \mathbf{K}-k^2\mathbf{Id} \\
	\end{pmatrix}
	\begin{pmatrix}
	\mathbf{i} \\
	\mathbf{E} \\
	\end{pmatrix}
	&=
	\begin{pmatrix}
	\frac{\mathbf{v}_\text{in-band}}{-j k \eta_0} \\
	\mathbf{0} \\
	\end{pmatrix}
	\\
	\label{eq:Sec-ImpedanceMatrix-DynamicalSystem3}
	\left[-k^2\begin{pmatrix}
	\mathbf{0} & \mathbf{0} \\
	\mathbf{0} & \mathbf{Id} \\
	\end{pmatrix}
	+\begin{pmatrix}
	\mathbf{0} & \mathbf{C} \\
	\mathbf{C}^T & \mathbf{K} \\
	\end{pmatrix}\right]
	\begin{pmatrix}
	\mathbf{i} \\
	\mathbf{E} \\
	\end{pmatrix}
	&=
	\begin{pmatrix}
	\frac{\mathbf{v}_\text{in-band}}{-j k \eta_0} \\
	\mathbf{0} \\
	\end{pmatrix}\text{,}
	\end{align}
\end{subequations}
where the state space for the {\color{black}\emph{electric field}} $\mathbf{E}$ in the microwave circuit (the electric field in the analysis domain, {\color{black}which is typically the unknown} for the CEM code), hidden to some degree in the impedance matrix transfer function, {\color{black}\emph{now shows up}. Note that \eqref{eq:Sec-ImpedanceMatrix-DynamicalSystem1} is simply another way to write down $\mathbf{Z}_\text{in-band}(k)$ in \eqref{eq:Sec-ImpedanceMatrix-SeriesDefinitionInBand} using matrix notations, where matrices $\mathbf{C}$ and $\mathbf{K}$ are detailed below. Getting from \eqref{eq:Sec-ImpedanceMatrix-DynamicalSystem1} to \eqref{eq:Sec-ImpedanceMatrix-DynamicalSystem2}, where the electric field $\mathbf{E}$ arises out of a sudden, may seem as if some kind of magic has been involved. This is not the case though. We play with advantage in CEM since we \emph{first} solve for the electric field $\mathbf{E}$ in the analysis domain and then compute the impedance matrix transfer function $\mathbf{Z}(k)$, cf. \cite{delaRubia2022EMCouplingMatrix}. As a matter of fact, in CEM, we normally go from \eqref{eq:Sec-ImpedanceMatrix-DynamicalSystem2} to \eqref{eq:Sec-ImpedanceMatrix-DynamicalSystem1}, but we can go backwards as well. In this work, we started off from the impedance matrix transfer function $\mathbf{Z}(k)$ and then arrived at the electric field state space $\mathbf{E}$. This has been so in an attempt to make the proposed approach more general.} From the discussion above, recall the pole-residue terms have valuable EM information. The coefficients $c_{np}$, included in the $\mathbf{C}^T$ matrix, stand for the electromagnetic coupling to port $p$ from each in-band eigenmode $n$. By the same token, the poles themselves, included in the $\mathbf{K}$ matrix as diagonal entries ($\mathbf{K}=\text{diag}\{k_1^2,\dots,k_N^2\}$), are the square of each eigenresonance, closely related to selfcoupling for each in-band eigenmode. {\color{black}Note that the dynamical system representation in \eqref{eq:Sec-ImpedanceMatrix-DynamicalSystem3} yields the in-band impedance matrix transfer function in \eqref{eq:Sec-ImpedanceMatrix-DynamicalSystem1}, $\mathbf{Z}_\text{in-band}(k)$, by simply eliminating the state variable $\mathbf{E}$ from \eqref{eq:Sec-ImpedanceMatrix-DynamicalSystem3} or from \eqref{eq:Sec-ImpedanceMatrix-DynamicalSystem2}.} It should be pointed out that the {\color{black}state space for the} electric field $\mathbf{E}$ in \eqref{eq:Sec-ImpedanceMatrix-DynamicalSystem} is expressed in the eigenmode basis, which indeed yields a diagonal $\mathbf{K}$ matrix. {\color{black}A similar methodology has been carried out earlier and the reader should not be overwhelmed, cf. \cite{amari2010theory,delaRubia2022EMCouplingMatrix}.} As a result, the following matrix, highlighted in \eqref{eq:Sec-ImpedanceMatrix-DynamicalSystem3},
\begin{equation}
	\label{eq:Sec-ImpedanceMatrix-CouplingMatrix}
	\begin{pmatrix}
	\mathbf{0} & \mathbf{C} \\
	\mathbf{C}^T & \mathbf{K} \\
	\end{pmatrix}\text{,}
\end{equation}
includes coupling information in transversal topology, cf.~\cite{Kurokawa,Conciauro,bekheit2008modeling,delaRubia2022EMCouplingMatrix}. We do not need to elaborate more on this. %matrix %(please, note the $k^2$ frequency behavior due to electromagnetics), matrix 
However, let us proceed the other way around. From a dynamical system point of view, \eqref{eq:Sec-ImpedanceMatrix-CouplingMatrix} is the matrix operator governing the second-order dynamical system in \eqref{eq:Sec-ImpedanceMatrix-DynamicalSystem3}. This matrix is responsible for the electric field dynamics and details the interactions among the dominant in-band eigenmodes in the EM circuit, thus providing valuable EM design information. This design information appears in the form of EM couplings among the in-band eigenmodes and ports in the coupled-resonator microwave circuit. As a result, this matrix \eqref{eq:Sec-ImpedanceMatrix-CouplingMatrix} stands for an electromagnetic coupling matrix.

Note that this matrix has nothing to do with the well-known circuit theory coupling matrix \cite{cameron2018microwave}. As a matter of fact, \eqref{eq:Sec-ImpedanceMatrix-DynamicalSystem3} is a second-order dynamical system (note the $-k^2$ frequency behavior in \eqref{eq:Sec-ImpedanceMatrix-DynamicalSystem3} due to \emph{electromagnetics}). This is not present in circuit theory, where a {\color{black}first-order} dynamical system arises, viz.
\begin{equation}
	\label{eq:Sec-ImpedanceMatrix-CircuitTheoryDynamicalSystem}
	\left[jk
	\begin{pmatrix}
	\mathbf{0} & \mathbf{0} \\
	\mathbf{0} & \mathbf{Id}\\
	\end{pmatrix}
	+		
	j\begin{pmatrix}
	\mathbf{0} & \mathbf{D} \\
	\mathbf{D}^T & \mathbf{M} \\
	\end{pmatrix}\right]
	\begin{pmatrix}
	\mathbf{i} \\
	\mathbf{e} \\
	\end{pmatrix}
	=
	\begin{pmatrix}
	\mathbf{v} \\
	\mathbf{0} 
	\end{pmatrix} \text{.}
\end{equation}
$\mathbf{e}$ is the vector with the {\color{black}state space (voltages at the \emph{circuit resonators})} and matrix
\begin{equation}
	\label{eq:Sec-ImpedanceMatrix-CircuitCouplingMatrix}
	\begin{pmatrix}
	\mathbf{0} & \mathbf{D} \\
	\mathbf{D}^T & \mathbf{M} \\
	\end{pmatrix}
\end{equation}
stands for the classical coupling matrix from circuit theory \cite{cameron2018microwave}. As a result, the \emph{circuit theory} impedance matrix $\mathbf{Z}(k)$ can be obtained by eliminating the internal {\color{black}state variable (voltages at the resonators)} $\mathbf{e}$ from \eqref{eq:Sec-ImpedanceMatrix-CircuitTheoryDynamicalSystem}. Thus,
\begin{equation}
	\label{eq:Sec-ImpedanceMatrix-CircuitImpedanceMatrix}
	\mathbf{v} = \mathbf{D} (j k \mathbf{Id}+j\mathbf{M})^{-1} \mathbf{D}^T \mathbf{i} = \mathbf{Z}(k) \mathbf{i}\text{.}
\end{equation}
	
{\color{black}Note that we stand upon electromagnetics and we would like to bridge this gap with circuit theory. As a result,} a \emph{first-order} Taylor approximation in the frequency variable $k$ can be carried out in the \emph{second-order} dynamical system in electromagnetics \eqref{eq:Sec-ImpedanceMatrix-DynamicalSystem} to account for the circuit theory behavior shown in \eqref{eq:Sec-ImpedanceMatrix-CircuitTheoryDynamicalSystem} and \eqref{eq:Sec-ImpedanceMatrix-CircuitImpedanceMatrix}. This approximation will {\color{black}be only} valid in a narrowband around the {\color{black}Taylor} expansion frequency. %{\color{black}Section \ref{Sec-NarrowbandAnalysis} will address this important point.}
\begin{remark}
	We will go through the details in Section \ref{Sec-NarrowbandAnalysis} and will discuss the degree to which the $k^2$ term shows up as well in the classical coupling matrix from circuit theory {\color{black}(after applying the low-pass to band-pass frequency transformation)}. In any case, within this new approach, it should be pointed out that we do not need to restrict ourselves to a narrowband analysis. The electromagnetic coupling matrix \eqref{eq:Sec-ImpedanceMatrix-CouplingMatrix} stands for the EM behavior of the microwave circuit in any frequency band of interest. Furthermore, this is so regardless of how tune or detune the EM device is {\color{black}since} we are solving Maxwell's equations with a CEM code. {\color{black}No fitting strategy can compete to this approach, specially for completely detuned EM circuits.} This will become more clear in Section \ref{Sec-NumericalResults}.	
\end{remark}

In order to get further physical insight in the band of interest by means of one single full-wave simulation of our coupled-resonator microwave circuit, we carry out a change of basis in the state space for the electric field $\mathbf{E}$. Instead of using the in-band eigenmodes as basis for the state space in the dynamical system \eqref{eq:Sec-ImpedanceMatrix-DynamicalSystem3} (which yields a diagonal $\mathbf{K}$ matrix by construction, i.e., a transversal topology coupling matrix arises, as mentioned above, cf.~\cite{cameron2018microwave}), we propose to use the \emph{local} resonators that appears in the microwave circuit as the basis in the state space for the electric field. These \emph{local} resonator fields are indeed a linear combination of the in-band \emph{global} eigenmodes. As a matter of fact, if the \emph{local} resonators vanish, the in-band \emph{global} eigenmodes do not show up. As a result, the EM couplings \emph{among all} local resonators and ports arise in the electromagnetic coupling matrix, detailing the actual internal state in the coupled-resonator EM circuit and providing valuable design information. Contrary to what has been previously done, no perfect null coupling coefficient (but weak) should be expected in the coupling matrix, since every local resonator couples to \emph{all} other local resonators to some degree \emph{in electromagnetics}. This representation enhances the physical insight provided in \cite{delaRubia2022EMCouplingMatrix}, where a \emph{specific} EM coupling topology is imposed by annihilating certain EM couplings, such as it is done in the classical approach \cite{cameron2018microwave}. Since our CEM code solves for the electric field $\mathbf{E}$ in the analysis domain, we have all the information to compute this more in-depth change of basis which identifies all EM couplings \emph{among} local resonators and ports. Section \ref{Sec-NumericalResults} discusses all these results. %we have all the ingredients to compute this change of basis. %we have full knowledge about computing this change of basis.
\begin{remark}
	{\color{black}The change of basis from local resonators to global eigenmodes is straightforward.} This is the change of basis that diagonalizes the $\mathbf{K}$ matrix, i.e., the eigenbasis in the linear map $\kappa(\cdot)$ associated to the $\mathbf{K}$ matrix, {\color{black}and can be obtained by simple eigendecomposition of the K matrix when the local resonator basis is used in this matrix. This guarantees the existence and uniqueness of a more in-depth change of basis, namely, the inverse change of basis}. As a matter of fact, we compute this inverse change of basis, which maps the global eigenmode basis onto the local resonator basis {\color{black}to get the local resonator EM coupling topology}. This is done with the help of the electric field $\mathbf{E}$ determined in the EM analysis. {\color{black}This change of basis is obtained with ease, but the procedure requires rather involved math. For the sake of understanding, we will not go through the details. %This margin is too narrow. 
	In any case, this software is completely available upon request.} %\emph{Cuius rei demonstrationem mirabilem sane detexi. Hanc marginis exiguitas non caperet.}} %We will skip the details as they are rather involved and they will be reported elsewhere. % We will not go through the details in this work since they are rather involved, and they will be reported elsewhere. %using the electric field $\mathbf{E}$ determined in the analysis domain. %The details are rather involved and will be reported elsewhere (patent pending). %The details are omitted due to space limitation.	
\end{remark}
\begin{remark}
	{\color{black} Finally, it should be pointed out that out-of-band eigenresonance contribution $\mathbf{Z}_\text{out-of-band}(k)$ is not dropped out at any point in this approach. $\mathbf{Z}_\text{out-of-band}(k)$ is a key factor in electormagnetics (in contrast to classical circuit theory) which stands for statics and higher-order modes contribution in the band of interest, and \emph{cannot}, therefore, \emph{be neglected}. We have already stressed this point out in \cite{delaRubia2022EMCouplingMatrix}, but we call it out once again for completeness. Note that other approaches based on classical circuit theory need to remove the so-called phase loading effect to satisfy the so-called compactness condition of the partial fraction expansion of the impedance matrix transfer function. As a matter of fact, this is the price to pay for neglecting the out-of-band eigenresonance contribution $\mathbf{Z}_\text{out-of-band}(k)$ in classical circuit approaches. This is not our case. All electromagnetic phenomena are accurately taken into account in our approach due to the fact that we are actually solving Maxwell's equations with a CEM code.}
\end{remark}
\section{Narrowband Analysis}
\label{Sec-NarrowbandAnalysis}
This new approach brings physical insight{\color{black}\cite{Bastioli2024,Molero2024}} and yet complements the available design information commonly used in coupled-resonator microwave circuits, namely, the coupling matrix in circuit theory. This classical approach is well-established. In spite of being an approximate model to account for electromagnetics, it is the preferential choice for design among microwave engineers{\color{black}\cite{cameron2018microwave,Bastioli2023,Tomassoni2024}}. In fact, there is no further alternative rather than costly parametric full-wave simulations.

In an attempt to bridge the gap between these two different methodologies (the classical coupling matrix and the electromagnetic coupling matrix), we carry out a \emph{narrowband approximation} in the electromagnetic coupling matrix approach. As a result, the second-order dynamical system arisen in electromagnetics \eqref{eq:Sec-ImpedanceMatrix-DynamicalSystem} is {\color{black}turned} into a {\color{black}first-order} dynamical system (of the same form as in \eqref{eq:Sec-ImpedanceMatrix-CircuitTheoryDynamicalSystem}, {\color{black}arisen in circuit theory}) by means of a proper frequency normalization and a first-order Taylor approximation around the expansion frequency. We are going through the details as this paves the way for some subtle variations that are currently under consideration to bring wideband behavior {\color{black}up} even at the circuit level description. This will allow us to show in the near future the electromagnetic coupling matrix as a circuit even for wideband analyses. So far, we stick ourselves to a \emph{narrowband analysis} to bridge the gap between both approaches.

First, the low-pass prototype frequency variable $\mathcal{K}$ (normalized frequency) is defined, which describes the \emph{classical} band-pass to low-pass transformation, viz.
\begin{equation}
\label{eq:Sec-NarrowbandAnalysis-LowPassPrototype}
%j\mathcal{K} = \frac{1}{FBW}(\frac{jk}{k_0}+\frac{k_0}{jk}).
\mathcal{K} = \frac{1}{\Delta}\left(\frac{k}{k_0}-\frac{k_0}{k}\right)\text{,}
% \kappa = \frac{1}{\Delta}\left(\frac{k}{k_0}-\frac{k_0}{k}\right)\text{.}
\end{equation}
%
%where $k_0 = \frac{\omega_0}{c}$, $\omega_0=\sqrt{\omega_1 \omega_2} $, and $\Delta = \frac{\omega_2-\omega_1}{\omega_0} $. 
where $k_0 = \frac{2\pi f_0}{c}$, $f_0=\sqrt{f_1 f_2} $, and $\Delta = \frac{f_2-f_1}{f_0} $. The \emph{narrowband analysis} is carried out within the frequency interval $[f_1, f_2]$. %Note that \eqref{eq:Sec-NarrowbandAnalysis-LowPassPrototype} requires $k_0\neq0$. 
Throughout this work, we use the term frequency for both the wavenumber $k$ and the frequency $f$ itself. %Another frequency normalization scheme is currently under investigation, and we intend to report these results in the near future. 
Second, we invert this mapping \eqref{eq:Sec-NarrowbandAnalysis-LowPassPrototype} to get the low-pass to band-pass transformation, namely,
\begin{equation}
\label{eq:Sec-NarrowbandAnalysis-kLowPassPrototype}
k = \frac{k_0}{2}\left(\mathcal{K} \Delta + \sqrt{\mathcal{K}^2 \Delta^2 + 4} \right)\text{,}
\end{equation}
and {\color{black}rewrite \eqref{eq:Sec-ImpedanceMatrix-DynamicalSystem3} to} put all the frequency dependence together in the dynamical system in \emph{electromagnetics} \eqref{eq:Sec-ImpedanceMatrix-DynamicalSystem}, viz.
\begin{equation}
	\label{eq:Sec-NarrowbandAnalysis-DynamicalSystem}
	\begin{aligned}
	\left[\begin{pmatrix}
	\mathbf{0} & \mathbf{0} \\
	\mathbf{0} & j k \mathbf{Id}+\frac{1}{j k} \mathbf{K} \\
	\end{pmatrix}
	+\begin{pmatrix}
	\mathbf{0} & \mathbf{C} \\
	\mathbf{C}^T & \mathbf{0} \\
	\end{pmatrix}\right]
	\begin{pmatrix}
	\mathbf{i} \\
	\mathbf{E} \\
	\end{pmatrix}
	&=
	\begin{pmatrix}
	\frac{\mathbf{v}_\text{in-band}}{-\eta_0} \\
	\mathbf{0} \\
	\end{pmatrix}\text{.}
	\end{aligned}
\end{equation}
%
%We assume that the \emph{eigenbasis} is used for the state space $\mathbf{E}$ such that the $\mathbf{K}$ matrix is \emph{diagonal}. 
Any EM coupling topology is taken into account so that the $\mathbf{K}$ matrix is, in general, a \emph{dense} matrix (no null coupling between resonators is assumed). Now that we have the physical frequency $k$ as a function of the normalized frequency $\mathcal{K}$, we plug in this relation \eqref{eq:Sec-NarrowbandAnalysis-kLowPassPrototype} into \eqref{eq:Sec-NarrowbandAnalysis-DynamicalSystem}. Thus,
\begin{equation}
	\label{eq:Sec-NarrowbandAnalysis-NormalizedFrequencyDynamicalSystem}
	\begin{aligned}
	\left[\begin{pmatrix}
	\mathbf{0} & \mathbf{0} \\
	\mathbf{0} & \mathbf{F}(j\mathcal{K}) \\
	\end{pmatrix}
	+\begin{pmatrix}
	\mathbf{0} & \mathbf{C} \\
	\mathbf{C}^T & \mathbf{0} \\
	\end{pmatrix}\right]
	\begin{pmatrix}
	\mathbf{i} \\
	\mathbf{E} \\
	\end{pmatrix}
	&=
	\begin{pmatrix}
	\frac{\mathbf{v}_\text{in-band}}{-\eta_0} \\
	\mathbf{0} \\
	\end{pmatrix}\text{,}
	\end{aligned}
\end{equation}
%
% where $\mathbf{F}(j\mathcal{K}) = \text{diag}\{ F_1(j\mathcal{K}),\dots,F_N(j\mathcal{K}) \}$ and
% %
% \begin{equation}
% 	\label{eq:Sec-NarrowbandAnalysis-NormalizedFrequencyFunction}
% 	\begin{aligned}
% 		F_n(j\mathcal{K}) &= \frac{j k_0}{2}\left(\mathcal{K} \Delta + \sqrt{\mathcal{K}^2 \Delta^2 + 4} \right) \\
% 			&+ \frac{2 k_n^2}{j k_0\left(\mathcal{K} \Delta + \sqrt{\mathcal{K}^2 \Delta^2 + 4} \right)} \text{,}
% 	\end{aligned}
% \end{equation}
%
where the matrix elements $F_{nm}(j\mathcal{K})$ in $\mathbf{F}(j\mathcal{K})$ are 
\begin{equation}
	\label{eq:Sec-NarrowbandAnalysis-NormalizedFrequencyFunction}
	\begin{aligned}
		F_{nm}(j\mathcal{K}) &= \frac{j k_0}{2}\left(\mathcal{K} \Delta + \sqrt{\mathcal{K}^2 \Delta^2 + 4} \right) \delta_{nm} \\
			&+ \frac{2}{j k_0\left(\mathcal{K} \Delta + \sqrt{\mathcal{K}^2 \Delta^2 + 4} \right)} K_{nm}\text{.}
	\end{aligned}
\end{equation}
$\delta_{nm}$ is the Kronecker delta and $K_{nm}$ are the matrix entries in $\mathbf{K}$, with $n,m = 1, \dots, N$. We carry out a first-order Taylor approximation in \eqref{eq:Sec-NarrowbandAnalysis-NormalizedFrequencyFunction} in the $\mathcal{K}$ variable centered at $\mathcal{K}=0$ (which implies using $k_0$ as expansion frequency in the physical frequency $k$). %Thus, $F_n(j\mathcal{K}) \approx F_n(j0) + F_n^\prime(j0) \mathcal{K}$, $n = 1, \dots, N$. 
Thus, $F_{nm}(j\mathcal{K}) \approx F_{nm}(j0) + F_{nm}^\prime(j0) \mathcal{K}$. %, $n,m = 1, \dots, N$. 
After this \emph{narrowband} approximation, the dynamical system \eqref{eq:Sec-NarrowbandAnalysis-NormalizedFrequencyDynamicalSystem} turns into
\begin{equation}
	\label{eq:Sec-NarrowbandAnalysis-NarrowbandNormalizedFrequencyDynamicalSystem}
	\begin{aligned}
	\left[j\mathcal{K}\begin{pmatrix}
	\mathbf{0} & \mathbf{0} \\
	\mathbf{0} & \frac{1}{j} \mathbf{F}^\prime(j0) \\
	\end{pmatrix}
	+ \begin{pmatrix}
	\mathbf{0} & \mathbf{C} \\
	\mathbf{C}^T & \mathbf{F}(j0) \\
	\end{pmatrix}\right]
	\begin{pmatrix}
	\mathbf{i} \\
	\mathbf{\tilde{E}} \\
	\end{pmatrix}
	&=
	\begin{pmatrix}
	\frac{\mathbf{v}_\text{in-band}}{-\eta_0} \\
	\mathbf{0} \\
	\end{pmatrix}\text{.}
	\end{aligned}
\end{equation}
{\color{black}Note that we} can no longer claim that the state space is the electric field $\mathbf{E}$, but an approximation $\mathbf{\tilde{E}}$. %Note that $\mathbf{F}^\prime(j\mathcal{K}) = \text{diag}\{ F_1^\prime(j\mathcal{K}),\dots,F_N^\prime(j\mathcal{K}) \}$. 
The procedure described thus far is similar to the one detailed in \cite{bekheit2008modeling}. However, only transversal coupling topology is taken into account in \cite{bekheit2008modeling}, where the $\mathbf{K}$ matrix is diagonal.

Further algebraic manipulations are carried out in \eqref{eq:Sec-NarrowbandAnalysis-NarrowbandNormalizedFrequencyDynamicalSystem} to arrive at the canonical circuit theory dynamical system (cf. \eqref{eq:Sec-ImpedanceMatrix-CircuitTheoryDynamicalSystem}), viz.
\begin{subequations}
	\label{eq:Sec-NarrowbandAnalysis-CircuitDynamicalSystem}
	\begin{align}
		\label{eq:Sec-NarrowbandAnalysis-CircuitDynamicalSystem1}
		\left[j\mathcal{K}
		\begin{pmatrix}
		\mathbf{0} & \mathbf{0} \\
		\mathbf{0} & \mathbf{Id}\\
		\end{pmatrix}
		+		
		j\begin{pmatrix}
		\mathbf{0} & \mathbf{D} \\
		\mathbf{D}^T & \mathbf{M} \\
		\end{pmatrix}\right]
		\begin{pmatrix}
		\mathbf{i} \\
		\mathbf{\tilde{e}} \\
		\end{pmatrix}
		&=
		\begin{pmatrix}
		\mathbf{v}_\text{in-band} \\
		\mathbf{0} 
		\end{pmatrix}	\\
		\label{eq:Sec-NarrowbandAnalysis-CircuitDynamicalSystem2}
		\mathbf{v}_\text{in-band} = \mathbf{D} (j \mathcal{K}\mathbf{Id}+j\mathbf{M})^{-1} \mathbf{D}^T \mathbf{i} &= \mathbf{Z}_\text{in-band}(\mathcal{K}) \mathbf{i}\text{,}
	\end{align}
\end{subequations}
where
\begin{equation}
	\label{eq:Sec-NarrobandAnalysis-Matrices}
	\begin{aligned}
		\mathbf{M} &=  [ \frac{1}{j} \mathbf{F}^\prime(j0) ]^{-\frac{1}{2}} [\frac{1}{j} \mathbf{F}(j0)] [ \frac{1}{j} \mathbf{F}^\prime(j0) ] ^{-\frac{1}{2}}\\
		\mathbf{D} &= \sqrt{\eta_0}\mathbf{C}[ \frac{1}{j} \mathbf{F}^\prime(j0)]^{-\frac{1}{2}}\\
		\mathbf{\tilde{e}} &= j \sqrt{\eta_0}[\frac{1}{j} \mathbf{F}^\prime(j0)]^{\frac{1}{2}} \mathbf{\tilde{E}} \text{.} 
	\end{aligned}
\end{equation}
% 
% Note that $\mathbf{F}(j0)$ and $\mathbf{F}^\prime(j0)$ are diagonal matrices.
% Note that $\mathbf{F}(j0)$ is a diagonal matrix.
An analogous procedure is carried out for the out-of-band term in \eqref{eq:Sec-ImpedanceMatrix-ImpendanceMatrix2}, $\mathbf{Z}_\text{out-of-band}(\mathcal{K})$. This term must not be neglected %to account for electromagnetics 
as it stands for the higher-order mode (and statics) loading in the band of analysis, which {\color{black}holds} \emph{only} in electromagnetics. {\color{black}Otherwise, this phase loading effect needs to be removed somehow.} %This term has been deliberately discarded in any circuit approximation to electromagnetics.
\begin{remark}
	The circuit theory coupling matrix shown in \eqref{eq:Sec-NarrowbandAnalysis-CircuitDynamicalSystem} is directly obtained from the electromagnetic coupling matrix in \eqref{eq:Sec-ImpedanceMatrix-DynamicalSystem}, resulting from Maxwell's equations, and the frequency interval %$[\omega_1, \omega_2]$, 
	$[f_1, f_2]$, which determines the narrowband analysis. It should be pointed out that the \emph{smaller} the fractional bandwidth~$\Delta$ (the narrower the band of interest), the \emph{more accurate} the approximation drawn upon circuit theory is to electromagnetics. %, which is completely described in \eqref{eq:Sec-ImpedanceMatrix-ImpendanceMatrix2} and~\eqref{eq:Sec-ImpedanceMatrix-DynamicalSystem}.
\end{remark}
\section{Numerical Results}
\label{Sec-NumericalResults}
In this section, we detail the narrowband electromagnetic coupling matrix (circuit representation) for several coupled-resonator microwave circuits, namely, a dual-mode cylindrical waveguide filter, an inline dielectric resonator filter, a combline diplexer and a dual-passband combline filter. Different tuning stages are {\color{black}detailed} in these examples {\color{black}to show the capabilities of the proposed approach}. Proper comparison to the classical circuit theory coupling matrix and what it is actually obtained in electromagnetics is discussed. All electromagnetic coupling matrices, as well as the narrowband electromagnetic coupling matrices, are available in \texttt{MATLAB} in reference \cite{delaRubiaMATLABexamples}. Note that the out-of-band term $\mathbf{Z}_\text{out-of-band}(\mathcal{K})$ is not detailed throughout this section, but it is never missed out in the analysis to properly describe any coupled-resonator EM circuit. {\color{black} Otherwise, the phase loading effect should be removed somehow, as we mentioned earlier.}%Thus, no port de-embedding is needed.

Our in-house C++ code for FEM simulations uses a second-order first family of N\'ed\'elec's elements \cite{Ned80, Ing06}, on meshes generated by \texttt{Gmsh} \cite{GeuR09}. However, as discussed in Section \ref{Sec-ImpedanceMatrix}, any CEM code which provides the impedance matrix transfer function of the EM device can be used. All computations were carried out on a workstation with {\color{black}two} 3.00-GHz Intel Xeon E5-2687W v4 {\color{black}processors} and 256-GB RAM.
\subsection{Dual-Mode Cylindrical Waveguide Filter}
\label{Sec-NumericalResults-Subsec-DualModeFilter}
\begin{figure}[tbp]
	\centering
	\includegraphics[width=\linewidth]{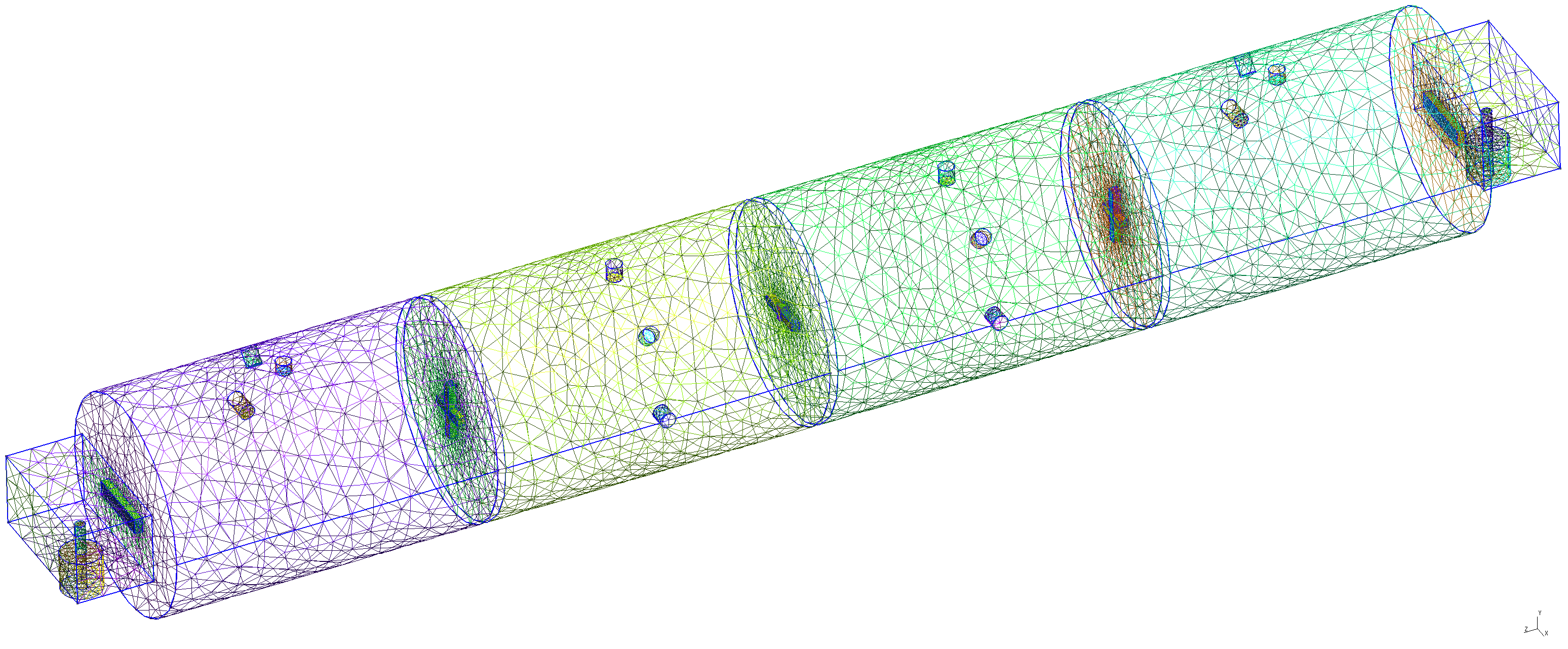}
	\caption{Dual-mode cylindrical waveguide filter.}
	\label{fig:Sec-NumericalResults-Subsec-DualModeFilter-FilterGeometry}
\end{figure}

An eight order cylindrical waveguide dual-mode filter is shown in Fig. \ref{fig:Sec-NumericalResults-Subsec-DualModeFilter-FilterGeometry}. The target filter response should provide 20 dB return losses in the 12.21-12.26 GHz passband and four transmission zeros in the following frequencies: 12.193, 12.201, 12.269 and 12.277 GHz. This target filter response is detailed by the classical coupling matrix in \eqref{eq:Sec-NumericalResults-Subsec-DualModeFilter-ClassicalCouplingMatrix}, which synthesis has been carried out with circuit theory \cite{Amari2000,cameron2018microwave}. %Cross-coupling between
\begin{table*}
	%	\small
	%   \tiny
	%	\footnotesize
	%	\scriptsize	
	%	\normalfont	
	\begin{equation}
	\label{eq:Sec-NumericalResults-Subsec-DualModeFilter-ClassicalCouplingMatrix}
	\begin{pmatrix}
 		0 &         0 &    0.9862 &         0 &         0 &         0 &         0 &         0 &         0 &         0\\
 		0 &         0 &         0 &         0 &         0 &         0 &         0 &         0 &         0 &    0.9862\\
   0.9862 &         0 &         0 &    0.8058 &         0 &   -0.1176 &         0 &         0 &         0 &         0\\
		0 &         0 &    0.8058 &         0 &    0.6626 &         0 &         0 &         0 &         0 &         0\\
		0 &         0 &         0 &    0.6626 &         0 &    0.5248 &         0 &         0 &         0 &         0\\
		0 &         0 &   -0.1176 &         0 &    0.5248 &         0 &    0.5307 &         0 &         0 &         0\\
		0 &         0 &         0 &         0 &         0 &    0.5307 &         0 &    0.4987 &         0 &   -0.2144\\
		0 &         0 &         0 &         0 &         0 &         0 &    0.4987 &         0 &    0.7249 &         0\\
		0 &         0 &         0 &         0 &         0 &         0 &         0 &    0.7249 &         0 &    0.7856\\
		0 &    0.9862 &         0 &         0 &         0 &         0 &   -0.2144 &         0 &    0.7856 &         0\\
	\end{pmatrix}
	%\text{.}
	\end{equation}
\end{table*}
\begin{table*}
	%	\small
	%   \tiny
	%	\footnotesize
	%	\scriptsize	
	%	\normalfont	
	\begin{equation}
	\label{eq:Sec-NumericalResults-Subsec-DualModeFilter-NarrowbandElectromagneticCouplingMatrix}
	\begin{pmatrix}
		0 &         0 &    1.0041 &   -0.1641 &    0.0000 &    0.0027 &    0.0011 &    0.0003 &    0.0002 &   -0.0000\\
		0 &         0 &    0.0000 &   -0.0002 &    0.0003 &   -0.0004 &    0.0072 &   -0.0001 &   -0.1604 &    1.0060\\
   1.0041 &    0.0000 &   -0.0272 &    0.8068 &    0.0968 &   -0.1007 &   -0.0059 &   -0.0013 &   -0.0007 &    0.0000\\
  -0.1641 &   -0.0002 &    0.8068 &   -0.2574 &    0.6462 &    0.0376 &    0.0018 &    0.0047 &    0.0022 &   -0.0008\\
   0.0000 &    0.0003 &    0.0968 &    0.6462 &   -0.0348 &    0.5274 &   -0.0179 &    0.0041 &   -0.0002 &   -0.0009\\
   0.0027 &   -0.0004 &   -0.1007 &    0.0376 &    0.5274 &    0.0200 &    0.5262 &    0.0105 &    0.0007 &    0.0008\\
   0.0011 &    0.0072 &   -0.0059 &    0.0018 &   -0.0179 &    0.5262 &   -0.0253 &    0.4882 &    0.0214 &   -0.2385\\
   0.0003 &   -0.0001 &   -0.0013 &    0.0047 &    0.0041 &    0.0105 &    0.4882 &    0.0217 &    0.7331 &    0.0996\\
   0.0002 &   -0.1604 &   -0.0007 &    0.0022 &   -0.0002 &    0.0007 &    0.0214 &    0.7331 &   -0.2426 &    0.7783\\
  -0.0000 &    1.0060 &    0.0000 &   -0.0008 &   -0.0009 &    0.0008 &   -0.2385 &    0.0996 &    0.7783 &   -0.0468\\
	\end{pmatrix}
	%\text{.}
	\end{equation}
\end{table*}

\begin{figure}[tbp]
	\centering
	\subfloat[]{
		\label{fig:Sec-NumericalResults-Subsec-DualModeFilter-PassbandResponse-ElectromagneticCouplingMatrix}
		\includegraphics[width=\linewidth]{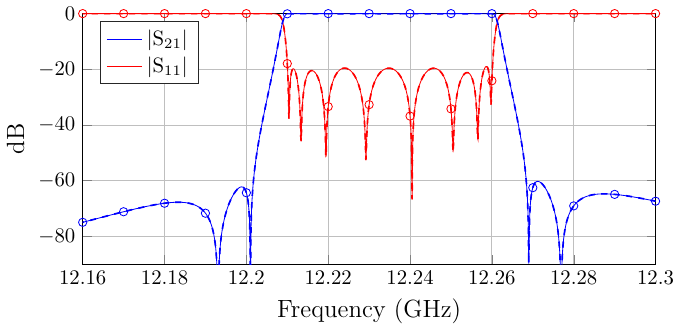}
	} \qquad \qquad \qquad
	\subfloat[]{
		\label{fig:Sec-NumericalResults-Subsec-DualModeFilter-PassbandResponse-ClassicalCouplingMatrix}
		\includegraphics[width=\linewidth]{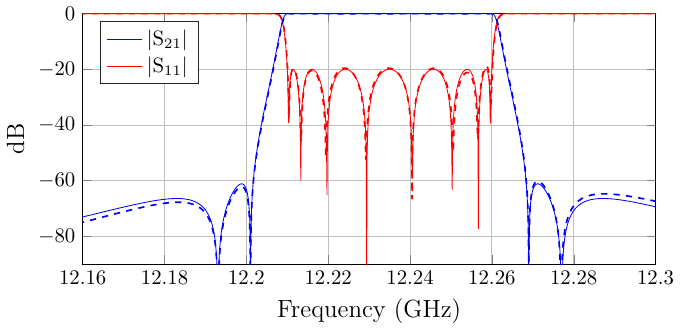}
	}
	\caption{Dual-mode cylindrical waveguide filter scattering parameter response. (a) Electromagnetic coupling matrix~[--]. FEM~[$\circ$]. Narrowband electromagnetic coupling matrix~[-~-]. (b) Classical coupling matrix~[--]. Narrowband electromagnetic coupling matrix~[-~-].  } 	
	\label{fig:Sec-NumericalResults-Subsec-DualModeFilter-PassbandResponse}
\end{figure}

An FEM discretization with 357,484 degrees of freedom is used. It takes 6~m 9~s to determine the impedance matrix transfer function and, as a result, the electromagnetic coupling matrix, within the 12.16-12.30 GHz band. Fig.~\ref{fig:Sec-NumericalResults-Subsec-DualModeFilter-PassbandResponse} compares the S parameter responses computed with FEM, the electromagnetic coupling matrix, the narrowband electromagnetic coupling matrix as well as the classical circuit theory coupling matrix \eqref{eq:Sec-NumericalResults-Subsec-DualModeFilter-ClassicalCouplingMatrix} obtained from synthesis. \emph{The EM circuit response was previously tuned using the proposed methodology}. The narrowband electromagnetic coupling matrix for this EM analysis is detailed in \eqref{eq:Sec-NumericalResults-Subsec-DualModeFilter-NarrowbandElectromagneticCouplingMatrix}.

Excellent agreement is obtained not only in the S parameter response with FEM and all the coupling matrix approaches, but also a reasonable concordance arises between the coupling matrices themselves, namely, the classical circuit theory coupling matrix \eqref{eq:Sec-NumericalResults-Subsec-DualModeFilter-ClassicalCouplingMatrix} and the narrowband electromagnetic coupling matrix \eqref{eq:Sec-NumericalResults-Subsec-DualModeFilter-NarrowbandElectromagneticCouplingMatrix}. As a matter of fact, we have made a great effort in this work to link both approaches. However, there are some phenomena that can {\color{black}{\emph{only}}} be captured within electromagnetics, such as {\color{black}parasitic and }leakage couplings among EM resonators. {\color{black}Look carefully at \eqref{eq:Sec-NumericalResults-Subsec-DualModeFilter-ClassicalCouplingMatrix} and \eqref{eq:Sec-NumericalResults-Subsec-DualModeFilter-NarrowbandElectromagneticCouplingMatrix}}. Even though the dominant coupling topology (the one intended in the classical coupling matrix \eqref{eq:Sec-NumericalResults-Subsec-DualModeFilter-ClassicalCouplingMatrix}) is resembled in electromagnetics, there is eventually no null (but weak or very weak) coupling among EM resonators. This is the result of electromagnetics, as it is very difficult to completely decouple neighboring resonators. This is not the case in circuit theory.

\begin{figure}[tbp]
	\centering
	\includegraphics[width=\linewidth]{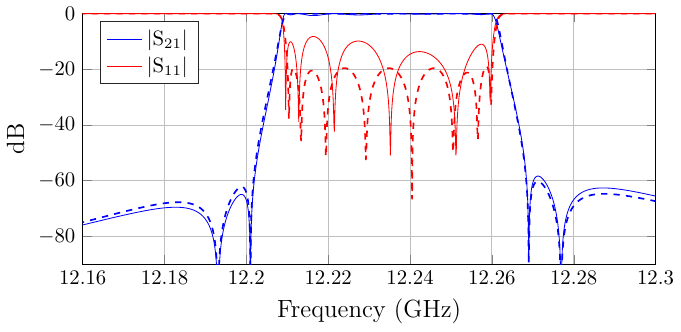}
	\caption{Dual-mode cylindrical waveguide filter scattering parameter response. Narrowband electromagnetic coupling matrix {\color{black}response} neglecting the out-of-band term~[--]. Narrowband electromagnetic coupling matrix {\color{black}response} taking into account the out-of-band term~[-~-]. } 	
	\label{fig:Sec-NumericalResults-Subsec-DualModeFilter-PassbandResponse-NoOutOfBand}
\end{figure}

Even though both coupling matrices \eqref{eq:Sec-NumericalResults-Subsec-DualModeFilter-ClassicalCouplingMatrix} and \eqref{eq:Sec-NumericalResults-Subsec-DualModeFilter-NarrowbandElectromagneticCouplingMatrix} look alike, we need to bear in mind that the narrowband electromagnetic coupling matrix \eqref{eq:Sec-NumericalResults-Subsec-DualModeFilter-NarrowbandElectromagneticCouplingMatrix} comes along with the out-of-band term $\mathbf{Z}_\text{out-of-band}(\mathcal{K})$. Contrary to what happens in circuit theory, higher-order mode and statics loading is present in the band of analysis in electromagnetics. If we neglect this term and plot the S parameter response out of the narrowband electromagnetic coupling matrix \eqref{eq:Sec-NumericalResults-Subsec-DualModeFilter-NarrowbandElectromagneticCouplingMatrix} itself, as if we were dealing with a classical circuit theory coupling matrix, we obtain the response depicted in Fig.~\ref{fig:Sec-NumericalResults-Subsec-DualModeFilter-PassbandResponse-NoOutOfBand}, which is clearly not electromagnetics. {\color{black}It is now evident the influence of the so-called phase loading effect, which we properly take into account in electromagnetics.}

% \begin{figure}[tbp]
% 	\centering
% 	% \input{figures/DualModeFilter_Sparameters_Response.tex}
% 	\includegraphics[width=\linewidth]{figures/DualModeFilter_Sparameters_Response.pdf}
% 	\caption{Dual-mode cylindrical waveguide filter scattering parameter response. (--)~Electromagnetic coupling matrix. ($\circ$) FEM. (-~-)~Narrowband electromagnetic coupling matrix. (-~-)~Classical coupling matrix.  } 	
% 	\label{fig:Sec-NumericalResults-Subsec-DualModeFilter-PassbandResponse}
% \end{figure}

%

\subsection{Inline Dielectric Resonator Filter}
\label{Sec-NumericalResults-Subsec-SnyderFilter}
\begin{figure}[tbp]
	\centering
	\includegraphics[width=\linewidth]{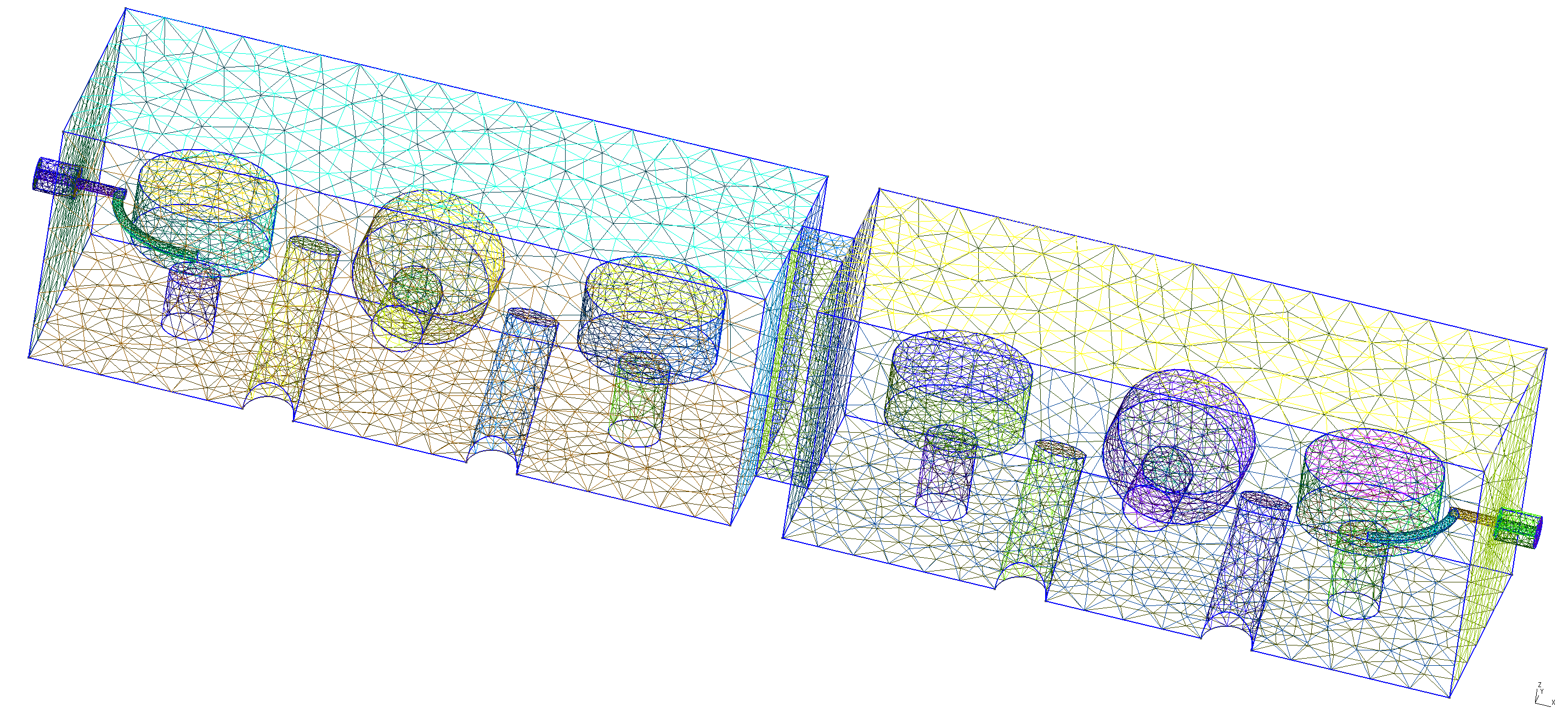}
	\caption{Inline dielectric resonator filter proposed in \cite{bastioli2012inline}.}
	\label{fig:Sec-NumericalResults-Subsec-SnyderFilter-FilterGeometry}
\end{figure}

A sixth order inline dielectric resonator filter is depicted in Fig. \ref{fig:Sec-NumericalResults-Subsec-SnyderFilter-FilterGeometry}. This filter was proposed in \cite{bastioli2012inline}. Two transmission zeros are generated by means of cross-coupling between nonadjacent dielectric resonators. The target filter response has a 20 dB return loss passband between 2.164 and 2.176 GHz. This time, a slightly detuned filter (not completely satisfying the passband specifications) is taken into account in the EM analysis. The resonators are increasingly numbered from left to right in Fig.~\ref{fig:Sec-NumericalResults-Subsec-SnyderFilter-FilterGeometry}.

\begin{figure}[tbp]
	\centering
	\subfloat[]{
		\label{fig:Sec-NumericalResults-Subsec-SnyderFilter-PassbandResponse-ElectromagneticCouplingMatrix}
		\includegraphics[width=\linewidth]{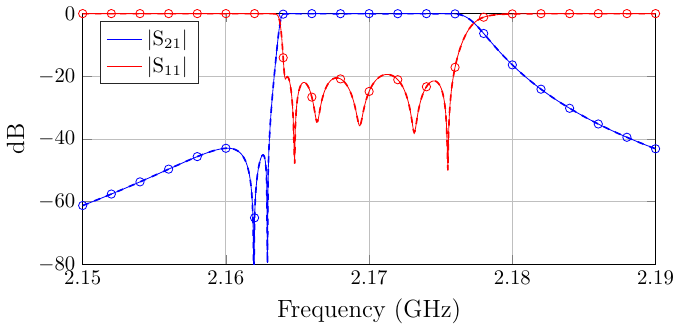}
	} \qquad \qquad \qquad
	\subfloat[]{
		\label{fig:Sec-NumericalResults-Subsec-SnyderFilter-PassbandResponse-ClassicalCouplingMatrix}
		\includegraphics[width=\linewidth]{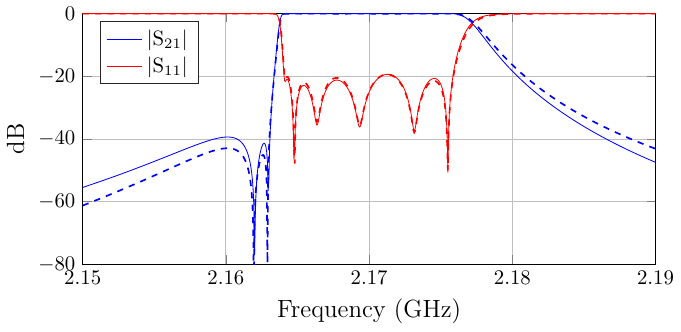}
	}
	\caption{Inline dielectric resonator filter scattering parameter response. (a) Electromagnetic coupling matrix~[--]. FEM~[$\circ$]. Narrowband electromagnetic coupling matrix~[-~-]. (b) Classical coupling matrix~[--]. Narrowband electromagnetic coupling matrix~[-~-].  } 	
	\label{fig:Sec-NumericalResults-Subsec-SnyderFilter-PassbandResponse}
\end{figure}

The circuit theory coupling matrix and the narrowband electromagnetic coupling matrix for this filter are detailed in \eqref{eq:Sec-NumericalResults-Subsec-SnyderFilter-ClassicalCouplingMatrix} and \eqref{eq:Sec-NumericalResults-Subsec-SnyderFilter-NarrowbandElectromagneticCouplingMatrix}, respectively. The classical coupling matrix is obtained by optimizing the location of the transmission zeros, reflection zeros and poles to the ones determined in the EM analysis, using the target electrical coupling matrix response from synthesis as starting point \cite{Amari2000,cameron2018microwave}. It takes 2~m 49~s to obtain the impedance matrix transfer function in the 2.15-2.19 GHz band. An FEM discretization with 229,468 degrees of freedom is used. The S parameter responses from {\color{black}FEM} analysis, electromagnetic coupling matrix and narrowband electromagnetic coupling matrix are depicted in Fig. \ref{fig:Sec-NumericalResults-Subsec-SnyderFilter-PassbandResponse-ElectromagneticCouplingMatrix}. Good agreement is found. In addition, Fig. \ref{fig:Sec-NumericalResults-Subsec-SnyderFilter-PassbandResponse-ClassicalCouplingMatrix} compares the S parameter responses from classical coupling matrix and the narrowband electromagnetic coupling matrix. By looking carefully at both coupling matrices \eqref{eq:Sec-NumericalResults-Subsec-SnyderFilter-ClassicalCouplingMatrix} and \eqref{eq:Sec-NumericalResults-Subsec-SnyderFilter-NarrowbandElectromagneticCouplingMatrix}, we are able to identify several similarities, including the dominant coupling topology. However, it is the narrowband electromagnetic coupling matrix the only one that details all the (leakage) EM couplings among resonators and ports, assuming no null couplings whatsoever. This brings further physical insight into the coupled-resonator EM circuit, showing clearly the internal EM state in the widely spread coupling matrix language. As a matter of fact, additional actions can now be locally taken into account to intensify or mitigate those dominant or unwanted couplings in the microwave filter.

\begin{figure}[tbp]
	\centering
	\includegraphics[width=\linewidth]{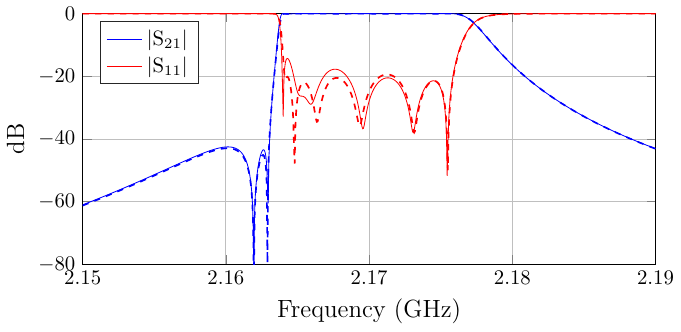}
	\caption{Inline dielectric resonator filter scattering parameter response. Narrowband electromagnetic coupling matrix increasing the iris width 0.2~mm~[--]. Narrowband electromagnetic coupling matrix from the initial state~[-~-]. } 	
	\label{fig:Sec-NumericalResults-Subsec-SnyderFilter-PassbandResponse-IrisChanged}
\end{figure}

In order to further show the physical insight this approach provides, we increase in 0.2 mm the width of the iris between resonators 3 and 4 and carry out the same EM analysis. The S parameter response is detailed in Fig. \ref{fig:Sec-NumericalResults-Subsec-SnyderFilter-PassbandResponse-IrisChanged}, where it can be seen that the filter response is further detuned. The narrowband electromagnetic coupling matrix is shown in \eqref{eq:Sec-NumericalResults-Subsec-SnyderFilter-NarrowbandElectromagneticCouplingMatrix-IrisChanged}. By comparing the narrowband electromagnetic coupling matrices \eqref{eq:Sec-NumericalResults-Subsec-SnyderFilter-NarrowbandElectromagneticCouplingMatrix} and \eqref{eq:Sec-NumericalResults-Subsec-SnyderFilter-NarrowbandElectromagneticCouplingMatrix-IrisChanged}, we identify that the largest change in the coupling {\color{black}coefficients take place in $M_{34}$ and $M_{43}$, which have} increased, as well as in $M_{33}$ and $M_{44}$, related to the resonant frequency of resonators 3 and 4, respectively, which have increased too, yielding a decrease in the resonant frequency. {\color{black}As a matter of fact}, increasing the width of the iris between resonators 3 and 4 should increase the coupling between these resonators but, at the same time, the resonators themselves are modified as well, increasing the size of their corresponding cavities. This {\color{black}results in} lower resonant frequencies for these resonators, which {\color{black}correspond} to an increase in the values of $M_{33}$ and $M_{44}$. All this shows that the narrowband electromagnetic coupling matrix is actually capturing the actual physical behavior in the coupled-resonator EM circuit.

\begin{table*}
	%	\small
	%   \tiny
	%	\footnotesize
	%	\scriptsize	
	%	\normalfont	
	\begin{equation}
	\label{eq:Sec-NumericalResults-Subsec-SnyderFilter-ClassicalCouplingMatrix}
	\begin{pmatrix}
		0 &        0 &   1.0160 &        0 &        0 &        0 &        0 &        0\\
		0 &        0 &        0 &        0 &        0 &        0 &        0 &   1.0098\\
   1.0160 &        0 &  -0.0023 &   0.7128 &  -0.4623 &        0 &        0 &        0\\
		0 &        0 &   0.7128 &   0.6036 &   0.4728 &        0 &        0 &        0\\
		0 &        0 &  -0.4623 &   0.4728 &  -0.1247 &   0.5819 &        0 &        0\\
		0 &        0 &        0 &        0 &   0.5819 &  -0.1203 &   0.3730 &  -0.5970\\
		0 &        0 &        0 &        0 &        0 &   0.3730 &   0.7885 &   0.6136\\
		0 &   1.0098 &        0 &        0 &        0 &  -0.5970 &   0.6136 &  -0.0177\\
	\end{pmatrix}
	%\text{.}
	\end{equation}
\end{table*}
%
% \begin{table*}
% 	%	\small
% 	%   \tiny
% 	%	\footnotesize
% 	%	\scriptsize	
% 	%	\normalfont	
% 	\begin{equation}
% 	\label{eq:Sec-NumericalResults-Subsec-SnyderFilter-ClassicalCouplingMatrix-IrisChanged}
% 	\begin{pmatrix}
% 		0 &        0 &   1.0126 &        0 &        0 &        0 &        0 &        0\\
% 		0 &        0 &        0 &        0 &        0 &        0 &        0 &   1.0097\\
%    1.0126 &        0 &  -0.0083 &   0.7113 &  -0.4615 &        0 &        0 &        0\\
% 		0 &        0 &   0.7113 &   0.6039 &   0.4732 &        0 &        0 &        0\\
% 		0 &        0 &  -0.4615 &   0.4732 &  -0.1104 &   0.5920 &        0 &        0\\
% 		0 &        0 &        0 &        0 &   0.5920 &  -0.1031 &   0.3719 &  -0.5981\\
% 		0 &        0 &        0 &        0 &        0 &   0.3719 &   0.7849 &   0.6128\\
% 		0 &   1.0097 &        0 &        0 &        0 &  -0.5981 &   0.6128 &  -0.0156\\
% 	\end{pmatrix}
% 	%\text{.}
% 	\end{equation}
% \end{table*}
%
\begin{table*}
	%	\small
	%   \tiny
	%	\footnotesize
	%	\scriptsize	
	%	\normalfont	
	\begin{equation}
	\label{eq:Sec-NumericalResults-Subsec-SnyderFilter-NarrowbandElectromagneticCouplingMatrix}
	\begin{pmatrix}
		0 &        0 &   1.0254 &  -0.1724 &   0.0196 &  -0.0010 &   0.0001 &  -0.0000\\
		0 &        0 &  -0.0001 &  -0.0004 &  -0.0011 &   0.0267 &  -0.1730 &   1.0192\\
   1.0254 &  -0.0001 &   0.0014 &   0.7997 &  -0.3850 &   0.0101 &  -0.0008 &   0.0003\\
  -0.1724 &  -0.0004 &   0.7997 &   0.3495 &   0.5498 &  -0.0100 &  -0.0003 &   0.0012\\
   0.0196 &  -0.0011 &  -0.3850 &   0.5498 &  -0.1232 &   0.5892 &  -0.0118 &   0.0134\\
  -0.0010 &   0.0267 &   0.0101 &  -0.0100 &   0.5892 &  -0.1197 &   0.4837 &  -0.5207\\
   0.0001 &  -0.1730 &  -0.0008 &  -0.0003 &  -0.0118 &   0.4837 &   0.5444 &   0.7433\\
  -0.0000 &   1.0192 &   0.0003 &   0.0012 &   0.0134 &  -0.5207 &   0.7433 &  -0.0233\\
	\end{pmatrix}
	%\text{.}
	\end{equation}
\end{table*}
\begin{table*}
	%	\small
	%   \tiny
	%	\footnotesize
	%	\scriptsize	
	%	\normalfont	
	\begin{equation}
	\label{eq:Sec-NumericalResults-Subsec-SnyderFilter-NarrowbandElectromagneticCouplingMatrix-IrisChanged}
	\begin{pmatrix}
		0 &        0 &   1.0226 &  -0.1716 &   0.0195 &  -0.0011 &   0.0001 &  -0.0000\\
		0 &        0 &  -0.0001 &  -0.0005 &  -0.0012 &   0.0267 &  -0.1731 &   1.0192\\
   1.0226 &  -0.0001 &  -0.0058 &   0.7994 &  -0.3845 &   0.0105 &  -0.0010 &   0.0004\\
  -0.1716 &  -0.0005 &   0.7994 &   0.3506 &   0.5496 &  -0.0099 &  -0.0004 &   0.0016\\
   0.0195 &  -0.0012 &  -0.3845 &   0.5496 &  -0.1076 &   0.6002 &  -0.0119 &   0.0137\\
  -0.0011 &   0.0267 &   0.0105 &  -0.0099 &   0.6002 &  -0.1022 &   0.4828 &  -0.5209\\
   0.0001 &  -0.1731 &  -0.0010 &  -0.0004 &  -0.0119 &   0.4828 &   0.5387 &   0.7427\\
  -0.0000 &   1.0192 &   0.0004 &   0.0016 &   0.0137 &  -0.5209 &   0.7427 &  -0.0213\\
	\end{pmatrix}
	%\text{.}
	\end{equation}
\end{table*}

{\color{black}Measurements of this device have been carried out in \cite{bastioli2012inline}. Good agreement with our simulation results can be found.}
\subsection{Combline Diplexer}
\label{Sec-NumericalResults-Subsec-ZhaoWuDiplexer}
\begin{figure}[tbp]
	\centering
	\includegraphics[width=1\linewidth,angle=0]{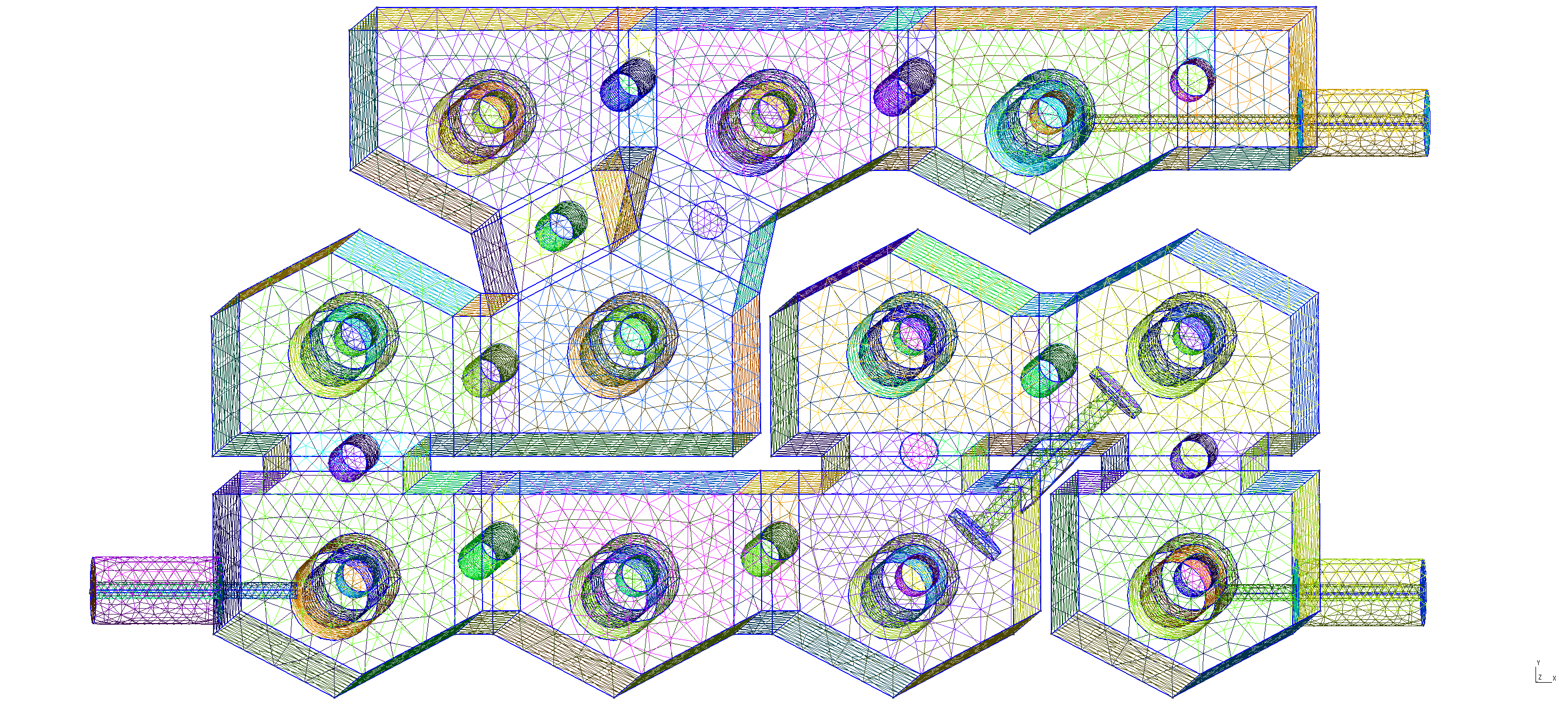}
	\caption{Combline diplexer designed in \cite{zhao2016model}.}
	\label{fig:Sec-NumericalResults-Subsec-ZhaoWuDiplexer-DiplexerGeometry}
\end{figure}

\begin{figure}[tbp]
	\centering
	\includegraphics[width=0.79\linewidth]{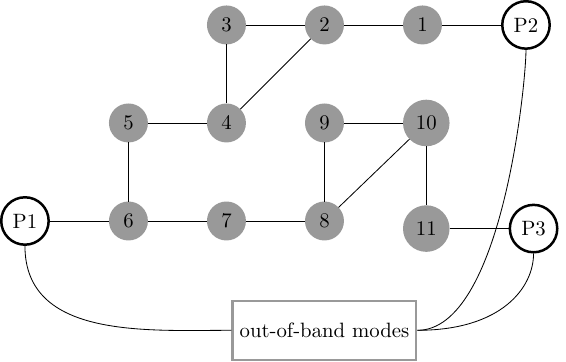}
	\caption{Intended coupling topology in the combline diplexer in Fig.~\ref{fig:Sec-NumericalResults-Subsec-ZhaoWuDiplexer-DiplexerGeometry}.}	
	\label{fig:Sec-NumericalResults-Subsec-ZhaoWuDiplexer-CouplingDiagram}
\end{figure}

\begin{figure}[tbp]
	\centering
	\includegraphics[width=\linewidth]{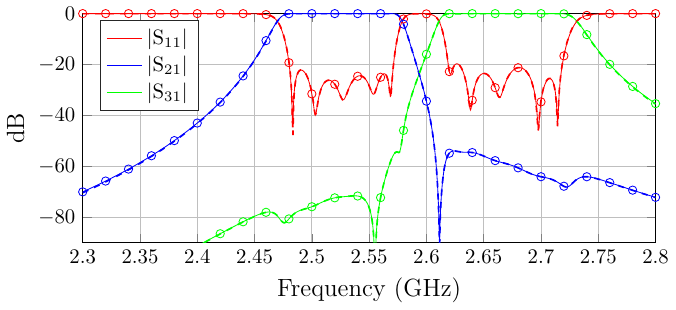}	
	\caption{Combline diplexer scattering parameter response. Electromagnetic coupling matrix~[--]. FEM~[$\circ$]. Narrowband electromagnetic coupling matrix~[-~-].} 	
	\label{fig:Sec-NumericalResults-Subsec-ZhaoWuDiplexer-DiplexerResponse}
\end{figure}

An eleventh order combline diplexer, proposed in \cite{zhao2016model}, is shown in Fig. \ref{fig:Sec-NumericalResults-Subsec-ZhaoWuDiplexer-DiplexerGeometry}. The EM analysis is carried out in the 2.3-2.8 GHz band. A reasonably tuned diplexer is taken into account and an FEM discretization with 272,188 degrees of freedom is used. The computation of the impedance matrix transfer function takes 6~m 2~s. The target diplexer passband channels are 2.478-2.568 GHz and 2.620-2.718 GHz, so the narrowband analysis is carried out in the 2.478-2.718 GHz band. The intended dominant coupling topology and resonator numbering are detailed in Fig. \ref{fig:Sec-NumericalResults-Subsec-ZhaoWuDiplexer-CouplingDiagram}, where the out-of-band contributions in electromagnetics are included as well. The S parameter responses from {\color{black}FEM} analysis, electromagnetic coupling matrix and narrowband electromagnetic coupling matrix are depicted in Fig. \ref{fig:Sec-NumericalResults-Subsec-ZhaoWuDiplexer-DiplexerResponse}. Good agreement is found. The narrowband electromagnetic coupling matrix for this diplexer is detailed in \eqref{eq:Sec-NumericalResults-Subsec-ZhaoWuDiplexer-NarrowbandElectromagneticCouplingMatrix}. The EM couplings resemble those of the intended dominant coupling topology, {\color{black}shown in Fig. \ref{fig:Sec-NumericalResults-Subsec-ZhaoWuDiplexer-CouplingDiagram}}. However, additional leakage EM couplings among resonators and ports are identified in this approach. This provides full control of the actual inner EM state in the diplexer. By means of one single full-wave simulation, the microwave engineer has now all the information to check the performance of the coupled-resonator EM circuit and carry out all the improvements needed to boost the EM response. This is done in a language the microwave engineer is familiar with, namely, the classical coupling matrix.

\begin{figure}[tbp]
	\centering
	\includegraphics[width=\linewidth]{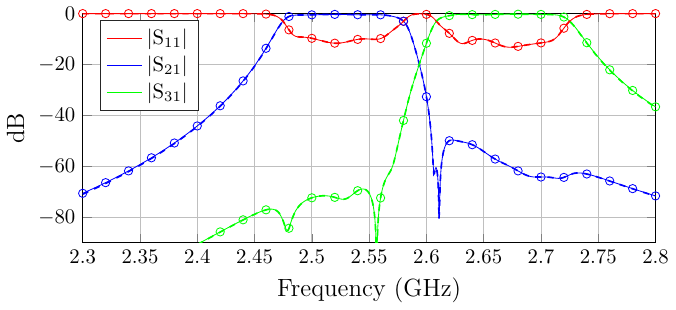}	
	\caption{Combline diplexer scattering parameter response changing the tuning screw depth in resonator 2 by -0.1 mm and in resonator 4 by +0.1 mm. Electromagnetic coupling matrix~[--]. FEM~[$\circ$]. Narrowband electromagnetic coupling matrix~[-~-].} 	
	\label{fig:Sec-NumericalResults-Subsec-ZhaoWuDiplexer-DiplexerResponse-Changed}
\end{figure}

Next, we change some tuning screws in the diplexer to verify that this approach actually provide physical insight capturing the EM behavior in the coupled-resonator EM circuit. The depths in the tuning screws in resonators 2 and 8 are modified by -0.1 mm (screw out) and +0.1 mm (screw in), respectively. The EM analysis is carried out in this new diplexer configuration. The S parameter responses in the FEM analysis, the electromagnetic coupling matrix and the narrowband coupling matrix are depicted in Fig. \ref{fig:Sec-NumericalResults-Subsec-ZhaoWuDiplexer-DiplexerResponse-Changed}, where we can see that both diplexer channels have been now detuned. The narrowband electromagnetic coupling matrix is shown in \eqref{eq:Sec-NumericalResults-Subsec-ZhaoWuDiplexer-NarrowbandElectromagneticCouplingMatrix-Changed}. By simple comparison in \eqref{eq:Sec-NumericalResults-Subsec-ZhaoWuDiplexer-NarrowbandElectromagneticCouplingMatrix} and \eqref{eq:Sec-NumericalResults-Subsec-ZhaoWuDiplexer-NarrowbandElectromagneticCouplingMatrix-Changed}, we can see that the coupling coefficients $M_{22}$ and $M_{88}$ are the ones which have been significantly modified in the expected direction. However, at the same time, we can identify that some neighboring couplings (those connected to nodes 2 and 8 in Fig. \ref{fig:Sec-NumericalResults-Subsec-ZhaoWuDiplexer-CouplingDiagram}) have been slightly modified as well, which is again expected in electromagnetics.
\begin{table*}
	%	\small
	  \tiny
	%   \footnotesize
	%	\scriptsize	
	%	\normalfont	
	\begin{equation}
	\label{eq:Sec-NumericalResults-Subsec-ZhaoWuDiplexer-NarrowbandElectromagneticCouplingMatrix}
	\begin{pmatrix}
		0 &        0 &        0 &   0.0000 &   0.0000 &   0.0000 &   0.0031 &   0.0830 &   1.1244 &   0.0841 &  -0.0109 &   0.0007 &  -0.0004 &  -0.0002\\
		0 &        0 &        0 &   0.6565 &   0.0175 &  -0.0007 &  -0.0000 &   0.0000 &  -0.0000 &   0.0000 &  -0.0000 &   0.0000 &  -0.0000 &   0.0000\\
		0 &        0 &        0 &  -0.0000 &  -0.0000 &  -0.0000 &   0.0000 &  -0.0000 &   0.0001 &  -0.0000 &   0.0016 &   0.0006 &   0.0213 &   0.6892\\
   0.0000 &   0.6565 &  -0.0000 &   0.6013 &   0.3478 &  -0.0247 &  -0.0046 &   0.0005 &  -0.0000 &   0.0000 &  -0.0000 &   0.0000 &  -0.0000 &   0.0000\\
   0.0000 &   0.0175 &  -0.0000 &   0.3478 &   0.6392 &   0.2256 &   0.1041 &  -0.0030 &  -0.0001 &  -0.0000 &   0.0000 &   0.0000 &  -0.0000 &  -0.0000\\
   0.0000 &  -0.0007 &  -0.0000 &  -0.0247 &   0.2256 &   0.4057 &   0.2246 &  -0.0066 &   0.0003 &   0.0000 &   0.0001 &  -0.0000 &  -0.0000 &  -0.0000\\
   0.0031 &  -0.0000 &   0.0000 &  -0.0046 &   0.1041 &   0.2246 &   0.6152 &   0.2872 &   0.0030 &   0.0022 &  -0.0000 &  -0.0000 &   0.0000 &   0.0000\\
   0.0830 &   0.0000 &  -0.0000 &   0.0005 &  -0.0030 &  -0.0066 &   0.2872 &   0.6152 &   0.5679 &   0.0662 &  -0.0099 &   0.0008 &  -0.0004 &  -0.0000\\
   1.1244 &  -0.0000 &   0.0001 &  -0.0000 &  -0.0001 &   0.0003 &   0.0030 &   0.5679 &  -0.4132 &   0.6322 &  -0.1028 &   0.0066 &  -0.0038 &   0.0007\\
   0.0841 &   0.0000 &  -0.0000 &   0.0000 &  -0.0000 &   0.0000 &   0.0022 &   0.0662 &   0.6322 &  -0.4152 &   0.2785 &   0.0116 &  -0.0040 &   0.0003\\
  -0.0109 &  -0.0000 &   0.0016 &  -0.0000 &   0.0000 &   0.0001 &  -0.0000 &  -0.0099 &  -0.1028 &   0.2785 &  -0.6354 &   0.2413 &  -0.0911 &   0.0070\\
   0.0007 &   0.0000 &   0.0006 &   0.0000 &   0.0000 &  -0.0000 &  -0.0000 &   0.0008 &   0.0066 &   0.0116 &   0.2413 &  -0.4414 &   0.2414 &  -0.0061\\
  -0.0004 &  -0.0000 &   0.0213 &  -0.0000 &  -0.0000 &  -0.0000 &   0.0000 &  -0.0004 &  -0.0038 &  -0.0040 &  -0.0911 &   0.2414 &  -0.6028 &   0.3546\\
  -0.0002 &   0.0000 &   0.6892 &   0.0000 &  -0.0000 &  -0.0000 &   0.0000 &  -0.0000 &   0.0007 &   0.0003 &   0.0070 &  -0.0061 &   0.3546 &  -0.7579\\	
	\end{pmatrix}
	%\text{.}
	\end{equation}
\end{table*}
\begin{table*}
	%	\small
	  \tiny
	%	\footnotesize
	%	\scriptsize	
	%	\normalfont	
	\begin{equation}
	\label{eq:Sec-NumericalResults-Subsec-ZhaoWuDiplexer-NarrowbandElectromagneticCouplingMatrix-Changed}
	\begin{pmatrix}
		0 &        0 &        0 &   0.0000 &   0.0000 &   0.0000 &   0.0031 &   0.0830 &   1.1243 &   0.0840 &  -0.0104 &   0.0006 &  -0.0004 &  -0.0002\\
		0 &        0 &        0 &   0.6567 &   0.0178 &  -0.0008 &  -0.0000 &   0.0000 &  -0.0000 &   0.0000 &  -0.0000 &   0.0000 &  -0.0000 &   0.0000\\
		0 &        0 &        0 &  -0.0000 &  -0.0000 &  -0.0000 &   0.0000 &  -0.0000 &   0.0001 &  -0.0000 &   0.0015 &   0.0006 &   0.0213 &   0.6892\\
   0.0000 &   0.6567 &  -0.0000 &   0.5987 &   0.3596 &  -0.0256 &  -0.0047 &   0.0005 &  -0.0000 &   0.0000 &  -0.0000 &   0.0000 &  -0.0000 &   0.0000\\
   0.0000 &   0.0178 &  -0.0000 &   0.3596 &   0.4939 &   0.2290 &   0.1035 &  -0.0030 &  -0.0001 &  -0.0000 &   0.0000 &   0.0000 &  -0.0000 &  -0.0000\\
   0.0000 &  -0.0008 &  -0.0000 &  -0.0256 &   0.2290 &   0.4055 &   0.2244 &  -0.0066 &   0.0003 &   0.0000 &   0.0001 &  -0.0000 &  -0.0000 &  -0.0000\\
   0.0031 &  -0.0000 &   0.0000 &  -0.0047 &   0.1035 &   0.2244 &   0.6249 &   0.2868 &   0.0030 &   0.0022 &  -0.0001 &  -0.0000 &   0.0000 &   0.0000\\
   0.0830 &   0.0000 &  -0.0000 &   0.0005 &  -0.0030 &  -0.0066 &   0.2868 &   0.6152 &   0.5679 &   0.0662 &  -0.0095 &   0.0007 &  -0.0004 &  -0.0000\\
   1.1243 &  -0.0000 &   0.0001 &  -0.0000 &  -0.0001 &   0.0003 &   0.0030 &   0.5679 &  -0.4126 &   0.6319 &  -0.0985 &   0.0062 &  -0.0036 &   0.0007\\
   0.0840 &   0.0000 &  -0.0000 &   0.0000 &  -0.0000 &   0.0000 &   0.0022 &   0.0662 &   0.6319 &  -0.4151 &   0.2812 &   0.0116 &  -0.0040 &   0.0002\\
  -0.0104 &  -0.0000 &   0.0015 &  -0.0000 &   0.0000 &   0.0001 &  -0.0001 &  -0.0095 &  -0.0985 &   0.2812 &  -0.5096 &   0.2405 &  -0.0918 &   0.0070\\
   0.0006 &   0.0000 &   0.0006 &   0.0000 &   0.0000 &  -0.0000 &  -0.0000 &   0.0007 &   0.0062 &   0.0116 &   0.2405 &  -0.4409 &   0.2413 &  -0.0060\\
  -0.0004 &  -0.0000 &   0.0213 &  -0.0000 &  -0.0000 &  -0.0000 &   0.0000 &  -0.0004 &  -0.0036 &  -0.0040 &  -0.0918 &   0.2413 &  -0.5995 &   0.3545\\
  -0.0002 &   0.0000 &   0.6892 &   0.0000 &  -0.0000 &  -0.0000 &   0.0000 &  -0.0000 &   0.0007 &   0.0002 &   0.0070 &  -0.0060 &   0.3545 &  -0.7611\\
	\end{pmatrix}
	%\text{.}
	\end{equation}
\end{table*}

{\color{black}Measurements of this diplexer have been detailed in \cite{zhao2016model}. Good agreement with our simulation results can be found in this reference.}
\subsection{Dual-Passband Combline Filter}
\label{Sec-NumericalResults-Subsec-ZhaoWuDualPassbandFilter}
\begin{figure}[tbp]
	\centering
	\includegraphics[width=1\linewidth,angle=0]{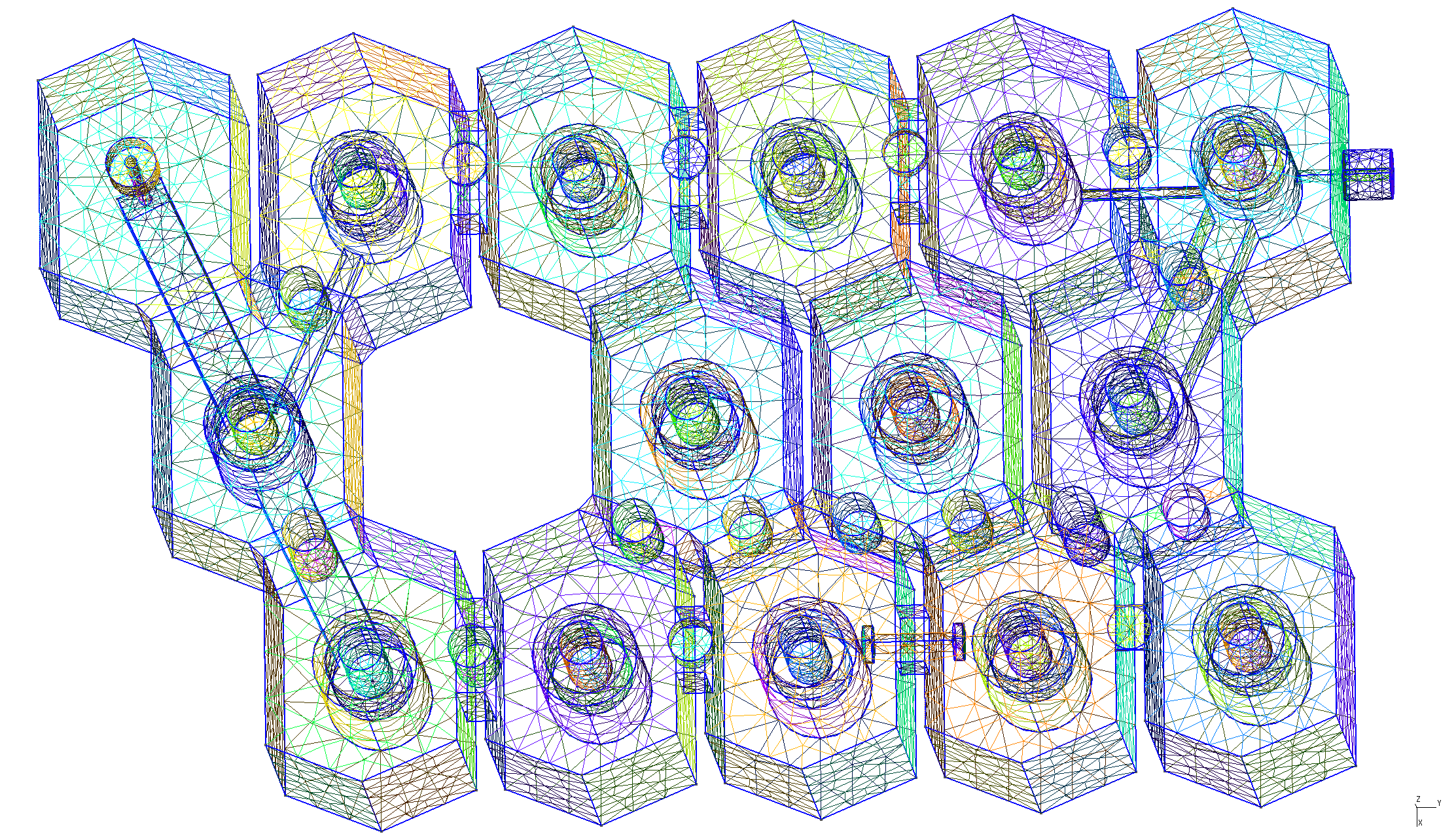}
	\caption{Dual-passband filter proposed in \cite{zhao2018circuit}.}
	\label{fig:Sec-NumericalResults-Subsec-ZhaoWuDualPassbandFilter-FilterGeometry}
\end{figure}

\begin{figure}[tbp]
	\centering
	\includegraphics[width=0.79\linewidth]{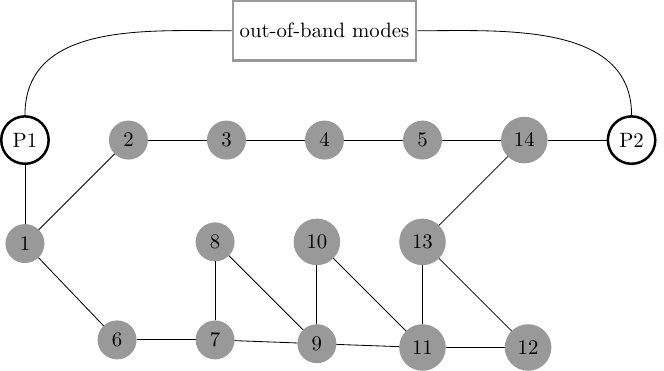}
	\caption{Intended coupling topology in the dual-passband filter in Fig.~\ref{fig:Sec-NumericalResults-Subsec-ZhaoWuDualPassbandFilter-FilterGeometry}.}	
	\label{fig:Sec-NumericalResults-Subsec-ZhaoWuDualPassbandFilter-CouplingDiagram}
\end{figure}

\begin{figure}[tbp]
	\centering
	\includegraphics[width=\linewidth]{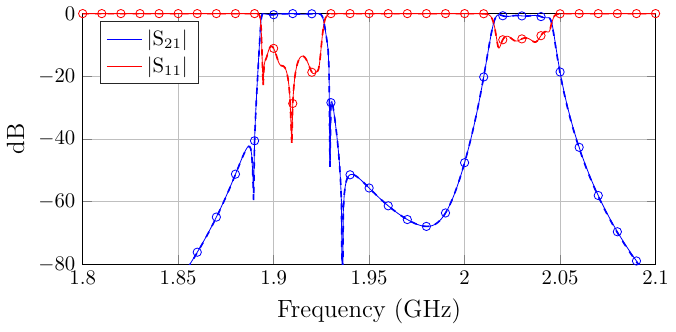}	
	\caption{Dual-passband filter scattering parameter response. Electromagnetic coupling matrix~[--]. FEM~[$\circ$]. Narrowband electromagnetic coupling matrix~[-~-].} 	
	\label{fig:Sec-NumericalResults-Subsec-ZhaoWuDualPassbandFilter-FilterResponse}
\end{figure}

A fourteenth order dual-passband combline filter, discussed in \cite{zhao2018circuit}, is detailed in Fig. \ref{fig:Sec-NumericalResults-Subsec-ZhaoWuDualPassbandFilter-FilterGeometry}. One possible realization is taken into account where we have deliberately studied a detuned filter response to show the performance of this technique even in this scenario. These detuned responses are of particular interest to the microwave engineer since these are the normal situations that need to be handled to carry out the final target design. The 1.8-2.1 GHz band is taken into account in the EM analysis. An FEM discretization with 338,988 degrees of freedom is used, and it takes 5~m 42~s to compute the impedance matrix transfer function. The nominal channels in the dual-passband filter are the 1.8845-1.9155 GHz and 2.009-2.026 GHz bands. As a result, the 1.8845-2.026 GHz band is taken into account for the narrowband analysis. The resonator numbering as well as the dominant intended coupling topology with the out-of-band contributions in electromagnetics are shown in Fig. \ref{fig:Sec-NumericalResults-Subsec-ZhaoWuDualPassbandFilter-CouplingDiagram}. The S parameter responses from {\color{black}FEM} analysis, electromagnetic coupling matrix and narrowband electromagnetic coupling matrix are detailed in Fig. \ref{fig:Sec-NumericalResults-Subsec-ZhaoWuDualPassbandFilter-FilterResponse}. Good agreement is found. The narrowband electromagnetic coupling matrix for this dual-passband filter is shown in \eqref{eq:Sec-NumericalResults-Subsec-ZhaoWuDualPassbandFilter-NarrowbandElectromagneticCouplingMatrix}. The intended coupling topology can be easily identified, but additional leakage couplings arise in the EM analysis.

\begin{figure}[tbp]
	\centering
	\includegraphics[width=\linewidth]{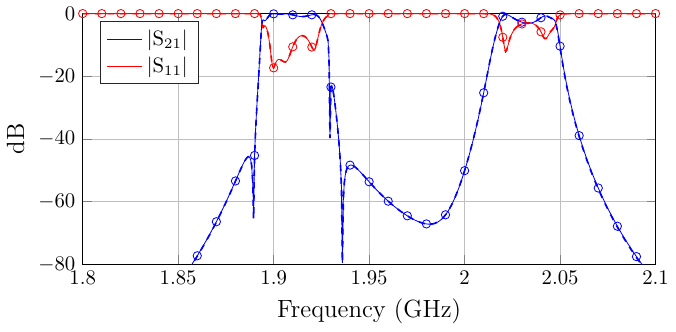}	
	\caption{Dual-passband filter scattering parameter response changing the depths by -0.1 mm in the tuning screw in resonator 2, in the tuning screw in resonator 7 and in the coupling screw between resonators 4 and 5. Electromagnetic coupling matrix~[--]. FEM~[$\circ$]. Narrowband electromagnetic coupling matrix~[-~-].} 	
	\label{fig:Sec-NumericalResults-Subsec-ZhaoWuDualPassbandFilter-FilterResponse-Changed}
\end{figure}

Next, we change some tuning and coupling screws in the dual-passband filter to check the performance of this approach even in this detuned scenario. The depths of the tuning screws in resonators 2 and 7, as well as the depth of the coupling screw between resonators 4 and 5 are decreased in 0.1 mm (screw out). The EM analysis is carried out in this new configuration. The S parameter responses in the FEM analysis, the electromagnetic coupling matrix and the narrowband coupling matrix are detailed in Fig. \ref{fig:Sec-NumericalResults-Subsec-ZhaoWuDualPassbandFilter-FilterResponse-Changed}, where we can see that both filter channels have been further detuned. The narrowband electromagnetic coupling matrix is shown in \eqref{eq:Sec-NumericalResults-Subsec-ZhaoWuDualPassbandFilter-NarrowbandElectromagneticCouplingMatrix-Changed}. All these results are completely available in \texttt{MATLAB} in reference \cite{delaRubiaMATLABexamples}. By comparing both narrowband electromagnetic coupling matrices \eqref{eq:Sec-NumericalResults-Subsec-ZhaoWuDualPassbandFilter-NarrowbandElectromagneticCouplingMatrix} and \eqref{eq:Sec-NumericalResults-Subsec-ZhaoWuDualPassbandFilter-NarrowbandElectromagneticCouplingMatrix-Changed}, we identify that the coupling coefficients $M_{22}$, $M_{77}$, {\color{black}$M_{45}$ and $M_{54}$} are the ones which have been significantly changed in the expected direction but, at the same time, we experience some changes of lower magnitude in the coupling coefficients $M_{44}$ and $M_{55}$ and some other coupling coefficients connected to resonators 2 and 7. All these changes are expected in electromagnetics and this approach can accurately predict their values, contrary to what has been previously done.

We should point out that all this valuable coupling information is ultimately used for design purposes. We do not go through the details in this work, but we will report some challenging designs in the near future using this new approach.
\begin{table*}
	%	\small
	  \tiny
	%   \footnotesize
	%	\scriptsize	
	%	\normalfont	
	\begin{equation}
	\label{eq:Sec-NumericalResults-Subsec-ZhaoWuDualPassbandFilter-NarrowbandElectromagneticCouplingMatrix}
	\begin{pmatrix}
		0 &        0 &   1.6203 &   0.0699 &   0.0010 &   0.0000 &   0.0000 &   0.2151 &  -0.0005 &   0.0000 &   0.0000 &   0.0000 &  -0.0000 &   0.0000 &   0.0000 &   0.0000\\
		0 &        0 &  -0.0000 &   0.0000 &  -0.0000 &   0.0002 &   0.0309 &   0.0000 &  -0.0000 &   0.0000 &  -0.0000 &  -0.0001 &   0.0003 &  -0.0000 &   0.0530 &  -1.4656\\
   1.6203 &  -0.0000 &  -1.2366 &   0.1895 &  -0.0014 &  -0.0000 &  -0.0000 &   0.7366 &  -0.0241 &   0.0001 &  -0.0001 &  -0.0000 &   0.0000 &  -0.0000 &  -0.0000 &  -0.0000\\
   0.0699 &   0.0000 &   0.1895 &  -1.0363 &   0.1079 &   0.0015 &   0.0000 &   0.0199 &  -0.0007 &   0.0000 &  -0.0000 &  -0.0000 &   0.0000 &  -0.0000 &  -0.0000 &   0.0000\\
   0.0010 &  -0.0000 &  -0.0014 &   0.1079 &  -1.0405 &   0.1079 &   0.0005 &  -0.0001 &   0.0000 &  -0.0000 &   0.0000 &  -0.0000 &  -0.0000 &   0.0000 &  -0.0000 &   0.0000\\
   0.0000 &   0.0002 &  -0.0000 &   0.0015 &   0.1079 &  -1.1081 &   0.1280 &  -0.0000 &   0.0000 &  -0.0000 &  -0.0000 &   0.0000 &  -0.0000 &   0.0000 &  -0.0002 &   0.0037\\
   0.0000 &   0.0309 &  -0.0000 &   0.0000 &   0.0005 &   0.1280 &  -1.0228 &  -0.0000 &   0.0000 &   0.0000 &  -0.0000 &   0.0000 &  -0.0001 &  -0.0000 &  -0.0093 &   0.2722\\
   0.2151 &   0.0000 &   0.7366 &   0.0199 &  -0.0001 &  -0.0000 &  -0.0000 &  -0.1221 &   0.2830 &   0.0011 &   0.0017 &   0.0000 &  -0.0000 &   0.0000 &   0.0000 &  -0.0000\\
  -0.0005 &  -0.0000 &  -0.0241 &  -0.0007 &   0.0000 &   0.0000 &   0.0000 &   0.2830 &   0.5826 &   0.1249 &   0.0519 &   0.0001 &  -0.0001 &  -0.0000 &  -0.0000 &  -0.0000\\
   0.0000 &   0.0000 &   0.0001 &   0.0000 &  -0.0000 &  -0.0000 &   0.0000 &   0.0011 &   0.1249 &   0.5317 &   0.1142 &   0.0045 &   0.0002 &  -0.0000 &  -0.0000 &  -0.0000\\
   0.0000 &  -0.0000 &  -0.0001 &  -0.0000 &   0.0000 &  -0.0000 &  -0.0000 &   0.0017 &   0.0519 &   0.1142 &   0.6170 &   0.0903 &  -0.0772 &   0.0002 &   0.0050 &   0.0004\\
   0.0000 &  -0.0001 &  -0.0000 &  -0.0000 &  -0.0000 &   0.0000 &   0.0000 &   0.0000 &   0.0001 &   0.0045 &   0.0903 &   0.7878 &   0.0998 &   0.0001 &   0.0067 &  -0.0003\\
  -0.0000 &   0.0003 &   0.0000 &   0.0000 &  -0.0000 &  -0.0000 &  -0.0001 &  -0.0000 &  -0.0001 &   0.0002 &  -0.0772 &   0.0998 &   0.5903 &   0.0103 &   0.2413 &   0.0243\\
   0.0000 &  -0.0000 &  -0.0000 &  -0.0000 &   0.0000 &   0.0000 &  -0.0000 &   0.0000 &  -0.0000 &  -0.0000 &   0.0002 &   0.0001 &   0.0103 &   0.3513 &   0.0903 &   0.0016\\
   0.0000 &   0.0530 &  -0.0000 &  -0.0000 &  -0.0000 &  -0.0002 &  -0.0093 &   0.0000 &  -0.0000 &  -0.0000 &   0.0050 &   0.0067 &   0.2413 &   0.0903 &  -0.1097 &   0.8163\\
   0.0000 &  -1.4656 &  -0.0000 &   0.0000 &   0.0000 &   0.0037 &   0.2722 &  -0.0000 &  -0.0000 &  -0.0000 &   0.0004 &  -0.0003 &   0.0243 &   0.0016 &   0.8163 &   0.7346\\
	\end{pmatrix}
	%\text{.}
	\end{equation}
\end{table*}
\begin{table*}
	%	\small
	  \tiny
	%	\footnotesize
	%	\scriptsize	
	%	\normalfont	
	\begin{equation}
	\label{eq:Sec-NumericalResults-Subsec-ZhaoWuDualPassbandFilter-NarrowbandElectromagneticCouplingMatrix-Changed}
	\begin{pmatrix}
		0 &        0 &   1.6204 &   0.0697 &   0.0010 &   0.0000 &   0.0000 &   0.2151 &  -0.0006 &   0.0000 &   0.0000 &   0.0000 &  -0.0000 &  -0.0000 &   0.0000 &   0.0000\\
		0 &        0 &   0.0000 &   0.0000 &  -0.0000 &   0.0002 &   0.0309 &  -0.0000 &  -0.0000 &   0.0000 &  -0.0000 &  -0.0001 &   0.0003 &  -0.0000 &   0.0530 &  -1.4657\\
   1.6204 &   0.0000 &  -1.2367 &   0.1913 &  -0.0014 &  -0.0000 &  -0.0000 &   0.7366 &  -0.0242 &   0.0001 &  -0.0001 &  -0.0000 &   0.0000 &  -0.0000 &  -0.0000 &  -0.0000\\
   0.0697 &   0.0000 &   0.1913 &  -1.1579 &   0.1079 &   0.0015 &   0.0000 &   0.0199 &  -0.0007 &   0.0000 &  -0.0000 &  -0.0000 &   0.0000 &   0.0000 &  -0.0000 &   0.0000\\
   0.0010 &  -0.0000 &  -0.0014 &   0.1079 &  -1.0407 &   0.1079 &   0.0005 &  -0.0001 &   0.0000 &  -0.0000 &   0.0000 &   0.0000 &  -0.0000 &   0.0000 &  -0.0000 &   0.0000\\
   0.0000 &   0.0002 &  -0.0000 &   0.0015 &   0.1079 &  -1.0996 &   0.1270 &  -0.0000 &   0.0000 &  -0.0000 &  -0.0000 &   0.0000 &  -0.0000 &   0.0000 &  -0.0002 &   0.0037\\
   0.0000 &   0.0309 &  -0.0000 &   0.0000 &   0.0005 &   0.1270 &  -1.0301 &  -0.0000 &   0.0000 &   0.0000 &  -0.0000 &   0.0000 &  -0.0001 &  -0.0000 &  -0.0093 &   0.2721\\
   0.2151 &  -0.0000 &   0.7366 &   0.0199 &  -0.0001 &  -0.0000 &  -0.0000 &  -0.1217 &   0.2821 &   0.0011 &   0.0017 &   0.0000 &  -0.0000 &   0.0000 &   0.0000 &  -0.0000\\
  -0.0006 &  -0.0000 &  -0.0242 &  -0.0007 &   0.0000 &   0.0000 &   0.0000 &   0.2821 &   0.4827 &   0.1257 &   0.0518 &   0.0001 &  -0.0001 &  -0.0000 &  -0.0000 &  -0.0000\\
   0.0000 &   0.0000 &   0.0001 &   0.0000 &  -0.0000 &  -0.0000 &   0.0000 &   0.0011 &   0.1257 &   0.5316 &   0.1142 &   0.0045 &   0.0002 &  -0.0000 &  -0.0000 &  -0.0000\\
   0.0000 &  -0.0000 &  -0.0001 &  -0.0000 &   0.0000 &  -0.0000 &  -0.0000 &   0.0017 &   0.0518 &   0.1142 &   0.6177 &   0.0903 &  -0.0773 &   0.0002 &   0.0050 &   0.0004\\
   0.0000 &  -0.0001 &  -0.0000 &  -0.0000 &   0.0000 &   0.0000 &   0.0000 &   0.0000 &   0.0001 &   0.0045 &   0.0903 &   0.7877 &   0.0998 &   0.0001 &   0.0067 &  -0.0003\\
  -0.0000 &   0.0003 &   0.0000 &   0.0000 &  -0.0000 &  -0.0000 &  -0.0001 &  -0.0000 &  -0.0001 &   0.0002 &  -0.0773 &   0.0998 &   0.5909 &   0.0103 &   0.2413 &   0.0243\\
  -0.0000 &  -0.0000 &  -0.0000 &   0.0000 &   0.0000 &   0.0000 &  -0.0000 &   0.0000 &  -0.0000 &  -0.0000 &   0.0002 &   0.0001 &   0.0103 &   0.3516 &   0.0903 &   0.0016\\
   0.0000 &   0.0530 &  -0.0000 &  -0.0000 &  -0.0000 &  -0.0002 &  -0.0093 &   0.0000 &  -0.0000 &  -0.0000 &   0.0050 &   0.0067 &   0.2413 &   0.0903 &  -0.1100 &   0.8163\\
   0.0000 &  -1.4657 &  -0.0000 &   0.0000 &   0.0000 &   0.0037 &   0.2721 &  -0.0000 &  -0.0000 &  -0.0000 &   0.0004 &  -0.0003 &   0.0243 &   0.0016 &   0.8163 &   0.7347\\
	\end{pmatrix}
	%\text{.}
	\end{equation}
\end{table*}

{\color{black}Measurements of this coupled-resonator circuit have been carried out in \cite{zhao2018circuit}. Reasonable agreement with our simulation results can be found.} %in that work.}
\section{Conclusions}
\label{Sec-Conclusions}
An enhanced electromagnetic coupling matrix detailing all EM couplings among resonators and ports in coupled-resonator EM circuits have been discussed in this work. Special effort has been made to show this novel EM coupling information in the same language as the one that it is commonly used by microwave engineers, namely, the classical circuit theory coupling matrix. The latter is traditionally used as a narrowband model for electromagnetics. In order to show the link between these two different approaches, a narrowband approximation in the electromagnetic coupling matrix has been introduced. Good agreement has been found between both approaches, i.e., the narrowband electromagnetic coupling matrix and the classical circuit theory coupling matrix. However, it is only via the EM analysis that all EM behavior has been captured into a coupling matrix theory in both, the electromagnetic coupling matrix and the narrowband electromagnetic coupling matrix. All leakage EM couplings have been identified in the EM circuit by means of one single EM analysis. As a result, valuable design information has been unleashed from the CEM code by means of the impedance matrix transfer function in the EM device.

Several coupled-resonator EM circuits such as a dual-mode filter, a dielectric resonator filter, a dual-passband filter and a diplexer have shown the capabilities %and possibilities 
of this new approach. 
\bibliographystyle{sty/IEEEtran}
\bibliography{bibliography/references}

%\bibliographystyle{../../../../sty/IEEEtran}
%\bibliography{../../../../bibtex/Bibliography/delarubia_IEEEabrv,../../../../bibtex/Bibliography/delarubia_IEEEbcpatabrv,../../../../bibtex/Bibliography/delarubia}

\end{document}